\documentclass[aps,preprint]{revtex4}
\usepackage{amsmath}
\usepackage{amssymb}
\usepackage{graphicx}
\usepackage{xcolor}

\newcommand{\func}[1]{\operatorname{#1}}

\begin{document}

\title{General hydrodynamic approach for a cold Bose gas}
\author{V.M. Pergamenshchik$^{1,2}$ }
\email{victorpergam@cft.edu.pl}
\date{\today }
\affiliation{$^{1}$Institute of Physics, National Academy of Sciences of Ukraine,
prospekt Nauky, 46, Kyiv 03039, Ukraine \\
$^{2}$Center for Theoretical Physics, Polish Academy of Sciences, Al. Lotnik\'{o}w 32/46, Warsaw, Poland \\
}

\begin{abstract}
The aim of this paper is to derive the hydrodynamics for a cold Bose gas
from the microscopic platform based on the many-body Schr\"{o}dinger
equation and general assumptions of the hydrodynamic approach (HA)
applicable to any dimension. We develop a general HA for a cold spatially
inhomogeneous Bose gas assuming two different temporal and spatial scales
and obtain the energy as a functional of both fast inner quantum mode and
slow macroscopic mode. The equations governing the fast and slow modes are
obtained from this functional by their independent variations. The fast mode
is the wave function in the stationary state at local density which can be
ground, excited with a nonzero atom momenta, or a superposition of more than
one states. The energy eigenvalue (or expectation value) of this local wave
function universally enters the hydrodynamic equation for the slow mode in
the form of the local chemical potential which incorporates the inner local
momentum. For zero inner momenta and particular choices of this eigenvalue
as a function of gas density, this equation reduces to the known equations
based on the local density approximation. If however the inner momenta are
nonzero, the equation includes the interaction between these momenta and the
slow mode velocity. Relation between this general HA and the standard local
density approximation is elaborated. Two effects of the local momenta and
their density dependence on the soliton solutions are demonstrated.
\end{abstract}

\pacs{}
\keywords{}
\maketitle


\section{Introduction}

The clouds of ultra cold Bose and Fermi gases in electromagnetic traps
naturally invoke the idea of a hydrodynamic-type description in terms of a
local order parameter, density, and velocity. To a great extent, the
successful development of this area of quantum physics has been based on the
hydrodynamic approach (HA) which intrinsically assumes the local density
approximation (LDA). However, at present the hydrodynamic-type approach to
quantum gases in terms of smoothly varying local quantities not only
continues to be used as one of the main theoretical tool, but also is a
subject of ongoing developments, modifications, and amendments. This work is
also motivated by the idea that not all capabilities of the HA and the
related LDA have already been fully explored and a further development is
still possible. There are (roughly) four main arguments that motivated this
work.

Firstly, there has been an ongoing chain of modifications and amendments to
the main mean field approach to quantum gases in the form of the
Gross-Pitaevskii equation (GPE). The GPE with a cubic nonlinearity was
proposed in the early 1960s \cite{Gross,Pitaevskii,Pit} for description of a
Bose-Einstein condensate. The idea underlying the GPE derivation is to
associate and replace the Heisenberg field operator with a classical mean
field order parameter which is assumed to vary slowly on distances of the
order of the interatomic force \cite{Pit}. The square of the modulus of this
order parameter was shown to represent gas density which in this way entered
the theory. As a result, via Madelung's ansatz, there have been obtained the
equivalent alternative hydrodynamic equations which formed the HA for a
superfluid. Although the GPE derivation does not resort to the LDA and the
equivalent superfluid HEs essentially express the mass and momentum
conservation \cite{Pit,Pethick}, the superfluid HA describes cold atomic gas
in terms of a local density and velocity, which has paved way to the natural
further development in the spirit of the conventional HA and the underlying
LDA. Different HAs have been developed and successfully applied to ultra
cold gases. But even before GPE appeared, Lee, Huang, and Yang \cite{LHY}
found what is now called a quantum correction to the GPE nonlinearity, which
appeared to be the lowest power of the gas density expansion; the GPE with
this correction is now known as the extended GPE (eGPE).

In 2000 Kolomeisky et al \cite{Kolom}, based on a renormalization group
analysis \cite{KolomPRB}, argued that in low dimensions strong atoms'
repulsion actually results in a quintic nonlinearity. The square of the
modulus of the mean field introduced in \cite{Kolom} is associated with the
local density and, as a result, the theory again becomes equivalent to
certain HA. It turned out, however \cite{restrictions}, that this theory
incorrectly predicts highly contrast interference patterns in the
one-dimensional split dynamics \cite{Girardeau2000} and shock wave formation
\cite{Peotta,Simons} in a strongly interacting Bose gas in the
Tonks-Gerardeau regime. Moreover, a similar situation was found to take
place also in a weakly interacting Bose gas in the GPE regime \cite%
{Peotta,Simons}. The reason is that the interference effects involve short
length oscillations which break the applicability of the LDA in general for
any interaction strength and, mathematically, the so-called quantum pressure
term is responsible for this problem \cite{Damski,Simons} (see Conclusion
section for a more discussion). Nevertheless, the HA explicitly or
indirectly exploiting the LDA remained attractive as it correctly describes
the situations with smooth density variations that often take place in the
experiment.

The intensive studies of a quasi-one-dimensional Bose gas in the early 2000s
brought about the idea, suggested by Dunjko, Lorent, and Olshanii \cite%
{Olshanii2001}, that in this geometry actual nonlinearity in the HA depends
on the local gas energy density \cite{Olshanii2001,Santos2002,Kim2003} and
can be taken from the Lieb-Liniger solution \cite{LL,L} (alternatively, it
is given by the equation of state obtained independently \cite{Astracharchik}%
). The advantage of such HA was recently demonstrated both for gas with a
short range repulsion \cite{Monopole,Pawl SciPost} and for a dipolar gas
\cite{PawlPRL}. In \cite{Monopole,PawlPRL,Pawl SciPost}, the
repulsion-induced nonlinearity in the one-dimensional hydrodynamic equation
was determined by the local energy density of the Lieb-Liniger gas. In these
papers, the starting point was the classical hydrodynamic equations for
local density and velocity. Then the quantum pressure term was formally
added to these equations because it is necessary to convert them to a Schr%
\"{o}dinger-type equation by means of the Madelung ansatz. The result of
such a conversion was dubbed Modified Nonlinear Schr\"{o}dinger Equation
(MNLSE) \cite{Monopole} and Lieb-Liniger GPE (LLGPE) \cite{Pawl SciPost}.
This conversion procedure was described in \cite{Monopole} as merely a
convenient numerical tool for simulating MNLSE instead of the hydrodynamic
equations with the remark that the hydrodynamic equations themselves should
be carefully examined in the future. Note that the conventional HA does not
account for the infinitely many integrals of motion of the Lieb-Liniger gas
which is adequately described by the recently developed generalized
hydrodynamics \cite{Bertini,Castro,Dubail}. Nevertheless, the standard HA
based on LDA, which is much simpler than the generalized hydrodynamics,
remains valid in many situations of interest and, in particular, for zero
temperature and in absence of velocity multivaluedness (see \cite{Review}).

We see that the modifications of the GPE depend on the specific system and
its dimension. The general HA in contrast is expected to predict a
hydrodynamic equation (HE) with certain term whose origin is universal and
only its specific form depends on the system and its geometry.\ In the
classical HEs such term is the pressure gradient which connects the
macroscopic motion with microscopic nature of the medium via the equation of
state. As such, the HA comprises two separate problems: first, finding the
pressure as a function of density and temperature for a given liquid with
its specific statistics, dimensions, molecular structure and so on, and,
second, solving the HE with this pressure. Such a HA, which is general in
the above sense, has not been developed for a generic cold Bose gas of any
origin and dimension.

Secondly, the fundamental relation between the microscopic molecular theory
and HA has been well established in the classical statistical physics. The
classical hydrodynamic equations can be derived from the many-body
distribution function by its reduction to the one-particle distribution and
then momentum integration. At the same time, the main idea of the derivation
of the HEs proposed for quantum gases in \cite%
{Gross,Pitaevskii,Pit,Olshanii2001,Santos2002,Kim2003,PawlPRL,Pawl SciPost}
is based on the essentially one-particle description and its hydrodynamic
form in the Madelung representation of the one-particle wave function.

Thirdly, the LDA, applied to a cold gas at zero temperature, assumes that
locally gas is in its ground state at the local density. However, a local
equilibrium at zero temperature implies only that the gas is in its
stationary state which is not necessarily ground state. For instance, a
stationary state can be an exited state with a nonzero local momentum or it
can be a superposition of different states. Such states can be incorporated
only on the level of many-particle descriptions. In particular, because in
these cases the total many-particle phase of system's wave function does not
have the form of the sum of individual atoms' phases and the one-particle
phase cannot be introduced.

The above arguments motivated us to derive the general HA to a cold Bose gas
which, on the one hand, would have a general form to justify the different
known equations on the common ground and, on the other hand, would provide
the connection with the many-body quantum mechanical description which is a
counterpart of the microscopic foundation of the classical HA. Our fourth
motivation argument is that, in principle, establishing such connection and
incorporating local non-ground states could reveal new effects and, in
particular, novel modifications of the known equations. In this paper this
program has been performed. As we have seen above, on the one hand, some
large scale descriptions of a cold quantum gas are explicitly HAs. On the
other hand, the large scale mean field description in terms of Scr\"{o}%
dnger-type or GPE-type equations via the Madelung transform also reduce to
certain equivalent HA. For this reason and for brevity, we will refer to
these approaches as known or standard HAs or known or standard HEs to
distinguish them from our HA and HE proposed in this paper.

The standard HA assumes that local small but still macroscopic subvolums are
in the thermodynamic equilibrium with some local temperature and density,
and that such subvolumes move as a whole with the local average velocity.
All these quantities change very little over distances of order of the
interatomic separation and thus are slow variables depending on the position
$X$ of the center of mass of the subvolumes. Thus, the LDA applied for a
quantum gas at finite temperature assumes the local thermodynamic
equilibrium \cite{T}. In the case of a quantum gas at zero temperature, we
assume that it is locally in a stationary state (\textit{ground or excited})
corresponding to the local atoms' density $n$, that it moves with a local
velocity $\mathbf{v},$ so that $n$ and $\mathbf{v}$ practically do not
change within a small local subvolumes and are slow functions of the
position of its center of mass and time. To this end we separate a slow
amplitude $A$ from the reduced one-body density matrix and find the HE for
this $A.$ The nonlinearity in this HE is specified by the density dependence
of the local chemical potential which in turn is determined by the eigen
values of the Schr\"{o}dinger equation for the locally homogeneous states.
It is this chemical potential that universally enters the HE while the
nonlinearity depends on the specific system and its dimension via the
specific density dependence of the local energy eigenvalues.

In the HA, only small macroscopic subsystems are in the locally stationary
states. This approach presupposes two different relaxation processes, a fast
microscopic and much slower hydrodynamic. At the first stage, small
subsystems equilibrate to their stationary states because their size $\Delta
V$ is much smaller than the total system size $V$ and their local
microscopic inner equilibrium is achieving much faster than the total one.
At the second, hydrodynamic stage, the density and velocity difference
between stationary subvolumes drives the slow macroscopic dynamics of the
whole system. We derive the energy as a functional of both fast mode and
slow mode. The fast mode is the wave function $\psi _{n}$ which is the
stationary state of the system Hamiltonian at the local density $n$ of the
subvolume. If the local stationary state has momentum excitations then the
local Hamiltonian in addition to the standard short range part also has the
term describing the interaction of the local momenta $\left\langle \mathbf{p}%
\right\rangle $ with the velocity of the slow mode. After the integrating
out the fast mode, we obtain the general HE only for slow mode $A$. This HE
contains the contribution from the local energy eigenvalue and the term $%
\left\langle \mathbf{p}\right\rangle \cdot \mathbf{\nabla }A$ describing
interaction between the fast and slow motions. The form of this HE is
universal and reduces the HA to the two universal problems: first finding
the stationary states $\psi $ for a given cold Bose gas at given density
and, second, solving the HE with their energy eigenvalues and finding $A.$
For $\left\langle \mathbf{p}\right\rangle =0$ and particular choice of the
energy eigenvalues as functions of gas density, the HE reduces to the known
equations. If however gas momenta are not due to its motion as a whole and
if these momenta depend on the gas density, the derived HA is the only tool
applicable in this case. Two examples of the influence of such momenta on
gray solitons are presented as an illustration to prompt possible related
effects.

In sec.II.A, we develop the general HA to a cold inhomogeneous Bose gas,
then in sec.II.B derive energy functional of both fast and slow modes and
impose local stationary states; in sec.II.C the HE is derived, discussed,
and its equivalent standard hydrodynamic form is obtained via the Madelung
transform. In sec.IIIA, we introduce the effective combined wave function of
both slow and fast modes, introduce its common coordinate description and,
in sec.III.B, present two examples of such function; then some important
general properties of our HA and its relation to the known GPE-type
equations are addressed in sec. III.C. In sec.IV, it is shown that, if local
momenta are nonzero, the periodic soliton train considered in \cite{Carr1}
can be made a propagating wave. In sec. V, we consider the effect of density
dependent local momentum on a soliton in a one-dimensional Lieb-Liniger gas,
and sec.VI. concludes the paper.

\section{From the microscopic to hydrodynamic description of a cold Bose gas}

\subsection{General consideration}

The system of our interest is an \textit{inhomogeneous} gas as its density
slowly varies in space. Consider such gas of $N~$cold boson atoms of mass $m$
in the volume $V,$ $\dim V=d$ ($d$ can be 1, 2, or 3)$.$ Let $x_{i}$ be a $d$%
-dimensional vector of the position of $i$-th atom, $x_{i}\in V,$ and $%
x=(x_{1},x_{2},...,x_{i},...,x_{N})$ be the $Nd$ dimensional vector
describing the total system in the $Nd$ dimensional volume $V\otimes
V\otimes V...=V^{N},$ $x\in V^{N};$ notation $x^{N-k}$ is used for the set
of coordinates of some $N-k$ atoms, $d^{N}x=dx_{1}...dx_{N}$ and $%
d^{N-k}x=dx_{k}...dx_{N}.$ The gas wave function $\psi (x,t)$ depends on $Nd$
spatial variables and is normalized on unity, $\left\langle \psi |\psi
\right\rangle =1.$ The gas Hamiltonian $\widehat{h}$ is the sum
\begin{equation}
\widehat{h}=\widehat{K}+\widehat{U}=-\frac{\hbar ^{2}}{2m}%
\sum_{i=1}^{N}\triangle _{i}+\sum_{i>j=1}^{N}U(x_{i}-x_{j}),  \label{hh}
\end{equation}%
where $\hbar $ is Plank's constant, $m$ is atom's mass, $\widehat{K}$ and $%
\widehat{U}$ are the operators of kinetic and potential energy, and $U$ is a
short range potential$;$ external and long range potentials will be
addressed later. The local density $n(y)$ is introduced through the reduced
one-body density matrix%
\begin{equation}
\rho (y,y^{\prime })=\int_{V^{N-1}}d^{N-1}x\psi ^{\ast }(y,x^{N-1})\psi
(y^{\prime },x^{N-1}).  \label{Ro}
\end{equation}%
The probability density $f(y)$ at $y$ is $f(y)=\rho (y,y)$ and the atom
density $n(y)=Nf(y)$.

To make a contact with subvolumes with different densities, for each
subvolume with the atom density $n$ we also introduce an \textit{auxiliary}
\textit{homogeneous} system $V_{n}$ of $N$ atoms at density $n$ which, by
this definition, has the volume $V_{n}=N/n,$ Fig.1. The idea is that, by
virtue of a short range interaction, the energy of a subvolume with density $%
n$ can be computed as a fraction of the energy of the associated auxiliary
system $V_{n}$ with the same density $n$. Consider the homogeneous system $%
V_{n}.$ Its wave function $\psi _{n}(x),$ which is normalized on unity, and
the energy $E_{n}$ are functions (functionals) of the density $n$ which is
indicated by a subscript $n.$\emph{\ }The system energy is%
\begin{equation}
E_{n}=\int_{V_{n}^{N}}d^{N}x\psi _{n}^{\ast }(x)\widehat{h}\psi _{n}(x).
\label{E0sr}
\end{equation}%
If the system is in a stationary state, then $E_{n}=N\varepsilon _{n}$ where
$E_{n}$ is the eigenvalue of the equation
\begin{equation}
\widehat{h}\psi _{n}=E_{n}\psi _{n}  \label{epsX}
\end{equation}%
and $\varepsilon _{n}$ is that per atom$.$ The energy of a short range
interaction is additive, i.e., the energy of a number of macroscopic
subvolumes is the sum of their energies. To see this explicitly in the case
of potential energy, the formula $U_{n}=\left\langle \psi _{n}|\widehat{U}%
|\psi _{n}\right\rangle $ can be presented in the following equivalent way
possible due to the symmetry with respect to atoms' coordinates $:$
\begin{equation}
U_{n}=\frac{N(N-1)}{2}\int_{V_{n}}dy\int_{V_{n}}dy^{\prime }U(y-y^{\prime
})G_{2}(y,y^{\prime }),  \label{Un}
\end{equation}%
where
\begin{equation}
G_{2}(y,y^{\prime })=\int_{V_{n}^{N-1}}d^{N-2}x\psi _{n}^{\ast }(y,y^{\prime
},x^{N-1})\psi _{n}(y,y^{\prime },x^{N-2})  \label{G2}
\end{equation}%
is the pair distribution. If the potential range is much shorter than the
size of the macroscopic subvolumes $\Delta V$ then
\begin{equation}
U_{n}=\frac{N(N-1)}{2}\sum_{\Delta V}\int_{\Delta V}dy\int_{\Delta
V}dy^{\prime }U(y-y^{\prime })G_{2}(y,y^{\prime }).  \label{Unn}
\end{equation}%
In particular, the potential energy $\Delta U_{n}$ of a macroscopic
subsystem with the number of atoms $\Delta N$ is the fraction of the total $%
U_{n}:$
\begin{equation}
\Delta U_{n}=\frac{\Delta N}{N}\left\langle \psi _{n}|\widehat{U}|\psi
_{n}\right\rangle .  \label{delVn}
\end{equation}

It is instructive to present similar relation for the kinetic energy in the
form that will help us to make a contact with the varying density. To this
end we separate the first or first two arguments of the wave function $\psi
_{n}(x)$: $\psi _{n}(y,x_{2},x_{3},..)=\psi _{n}(y,x^{N-1}),\psi
_{n}(y,y^{\prime },x_{3},...)=\psi _{n}(y,y^{\prime },x^{N-2})$ where
symbols $y$ and $y^{\prime }$ stand for the $d$ dimensional vector of
position of a single atom, a point in $V$. Then the kinetic energy $K_{n}=$ $%
\left\langle \psi _{n}|\widehat{K}|\psi _{n}\right\rangle $ can be presented
in terms of the reduced density matrix $\rho _{n}(y,y^{\prime })$ of the
system $V_{n}:$%
\begin{equation}
K_{n}=-\frac{\hbar ^{2}N}{2m}\int_{V_{n}}dy\int_{V_{n}}dy^{\prime }\delta
(y-y^{\prime })\triangle _{y^{\prime }}\rho _{n}(y,y^{\prime }),  \label{Kn}
\end{equation}%
where
\begin{equation}
\rho _{n}(y,y^{\prime })=\int_{V_{n}^{N-1}}d^{N-1}x\psi _{n}^{\ast
}(y,x^{N-1})\psi _{n}(y^{\prime },x^{N-1})  \label{ro}
\end{equation}%
is the reduced one-body density matrix in a homogeneous gas with the density
$n$. It follows that the kinetic energy $\Delta K_{n}$ of a macroscopic
susbsystem of $\Delta N$ atoms is
\begin{eqnarray}
\Delta K_{n} &=&-\frac{\hbar ^{2}N}{2m}\int_{\Delta V}dy\int_{\Delta
V}dy^{\prime }\delta (y-y^{\prime })\triangle _{y^{\prime }}\rho
_{n}(y,y^{\prime })  \label{delKn} \\
&=&\frac{\Delta N}{N}K_{n}.  \notag
\end{eqnarray}%
The probability density $f_{n}(y)=\rho _{n}(y,y)$ and atom density $n=Nf_{n}$
in a homogeneous system are constant. Equations (\ref{Kn})-(\ref{delKn})
will be used to introduce slow density variations and obtain the energy
functional in which the slow and fast modes are separated.

Consider now our inhomogeneous system with slowly modulated $f(y)$ and $%
n(y). $ We divide the volume $V$ into small macroscopic subvolumes of equal
volume $\Delta V\ll V$ with the linear size $\Delta V^{1/d}$ much larger
than the average interatomic separation $n^{-1/d},$ so that the number of
atoms $n\Delta V$ in $\Delta V$ is large$.$ Let $y=X$ be the position of the
center of $\Delta V_{X},\sum_{X}\Delta V_{X}=V$, Fig.1.
\begin{figure}[tph]
\centering
\includegraphics[width=\textwidth]{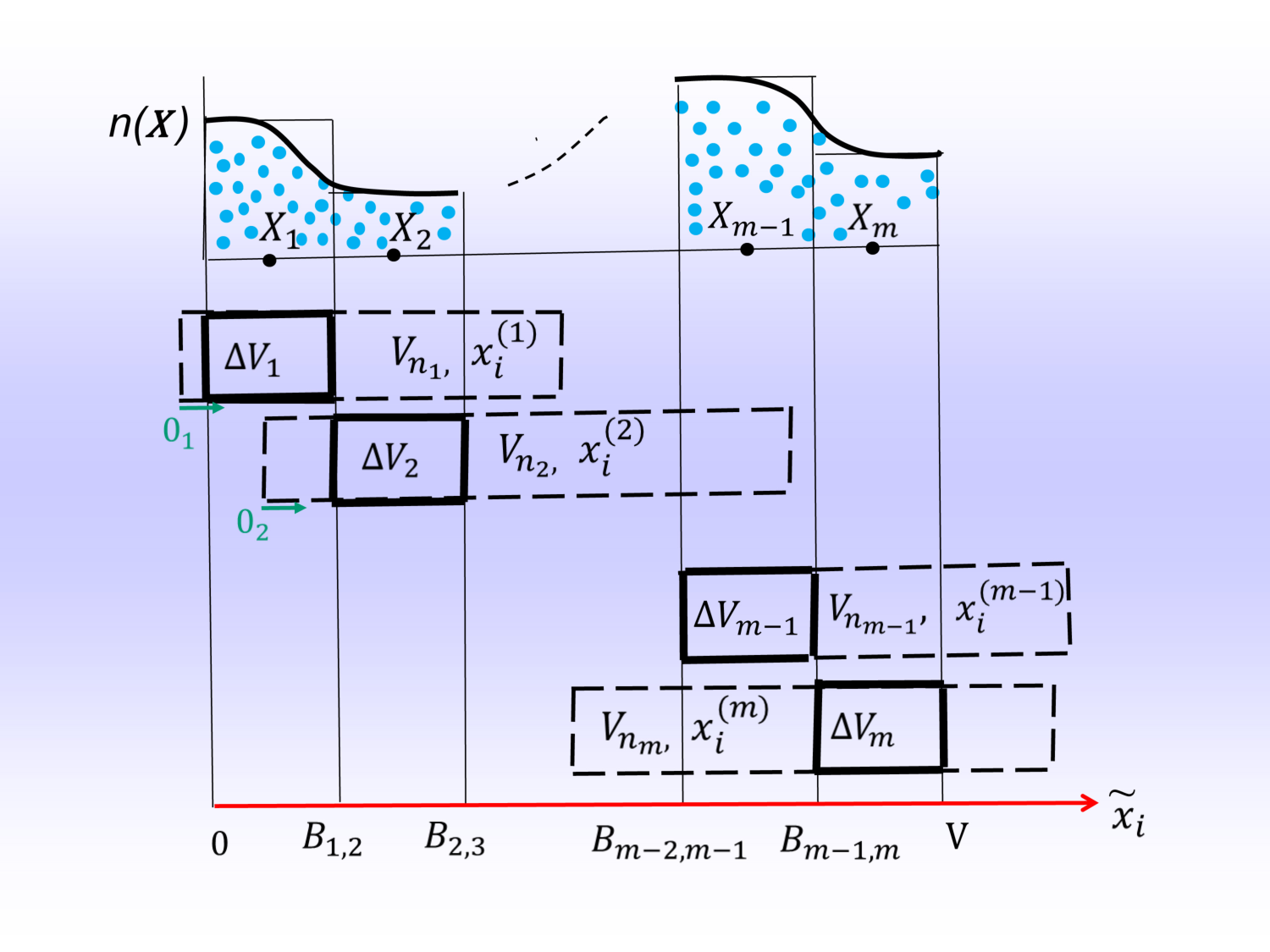}
\caption{One-dimensional sketch of the main construction of our approach. An
inhomogeneous Bose gas divided into $m$ quasi-homogeneous subvolumes $\Delta
V_{k}$ with the atom density $n_{k}$ and which are centered at $X_{k}$. At
the first stage, each $\Delta V_{k}$ (bold rectangles) is considered a part
of an auxiliary homogeneous gas of $N$ atoms (dashed rectangles), which has
the density $n_{k}$ and volume $V_{n_{k}}=N/n_{k};$ $V_{n_{k}}$ is larger
for lower density $n_{k}$. The $d$-dimensional vector of the coordinate $%
x_{i}^{(k)}$ of $i$ th atom, $i=1,2,...N,$ is defined in the entire
auxiliary volume $V_{n_{k}}$ relative to some reference frame $O_{k}.$ The
onsets of $O_{k}$ in different $V_{n_{k}}$ are not correlated; for instance,
they can be connected to the left ends of $V_{n_{k}}$ as shown by the arrows
from $O_{1}$ and $O_{2}.$ Thus, the atoms' coordinates $x_{i}^{(k)}$ in
different $V_{n_{k}}$ are not mutually adjusted. But, at the second,
hydrodynamic stage, the atoms are described only in the actual volume $V$
which requires the vectors $\widetilde{x}_{i}\in V,$ continuously
extrapolating local $x_{i}^{(k)}$ between different $\Delta V_{k}.$ This $%
\widetilde{x}_{i}$ can be obtained by synchronizing $x_{i}^{(k)}$in
different $\Delta V_{k}$ by taking the common onset $O$ for all $O_{k}$, as
shown by the bottom arrow. Then at the boundary $B_{1,2}$ between $\Delta
V_{1}$ and $\Delta V_{2},$ $x_{i}^{(1)}=x_{i}^{(2)},..,$ and so on. As a
result, defining $\widetilde{x}_{i}$ to be equal to $x_{i}^{(k)}$ within $%
\Delta V_{k}$ gives the desired continuous position vector of $i$ th atom in
the actual volume $V.$}
\end{figure}
We will indicate quantities related to $\Delta V_{X}$ centered at $X$ by a
subscript $X,$ i.e., if $y\in \Delta V_{X}$ then $%
f(y)=f_{X}(y),n(y)=n_{X}(y) $; these functions computed at $y=X$ will be
denoted by the same symbols but without the argument, i.e., $%
f_{X}(X)=f_{X},n_{X}(X)=n_{X}.$ Clearly, at different $X$ the functions $%
f_{X}$ and $n_{X}$ can be very different, but their dependence on $X$ is
slow and the total energy is the sum of energies of the subvolumes $\Delta
V_{X}:$
\begin{equation}
E=\sum_{X}\Delta E_{X}=\sum_{X}(\Delta K_{X}+\Delta U_{X}).  \label{Esum}
\end{equation}%
As for $y\in \Delta V_{X}$ one has $f_{X}(y)\simeq f_{X},n_{X}(y)\simeq
n_{X} $, $\sum_{X}f_{X}\Delta V_{X}=1,$ the energy terms without derivatives
can be taken at $y=X$ which accounts for their dependence on the
coarse-grained coordinate $X$ since their variations within $\Delta V_{X}$
can be neglected. However, differentiation of $\rho (y,y^{\prime })$ in $%
\Delta K_{n}$ (\ref{delKn}) can result in the kinetic energy terms due to a
slow but finite density variation $n_{X}(y)-n_{X}$ even within each $\Delta
V_{X}. $ To obtain these terms, in Appendix A we separate from the density $%
n_{X}(y) $ a slow amplitude $A_{X}(y)$ which accounts for the density
deviation relative to $n_{X}$ within $\Delta V_{X}.$ Namely, for each $X$
and atom's coordinates $y$ and $y^{\prime }$ \textit{within} $\Delta V_{X},$
in the full local one-body density matrix $\rho _{n}(y,y^{\prime })$ in $%
\Delta V_{X}$ we separate the factor $A_{X}^{\ast }(y)A_{X}(y^{\prime })$
due to the variation $n_{X}(y)-n_{X}$ and obtain the following relation:
\begin{eqnarray}
\rho _{n}(y,y^{\prime }) &=&\frac{A_{X}(y)^{\ast }A_{X}(y^{\prime })}{f_{X}}%
\rho _{X}(y,y^{\prime }),  \label{f=af} \\
y,y^{\prime } &\in &\Delta V_{X},  \notag
\end{eqnarray}%
where $f_{X}=A_{X}{}^{\ast }(X)A_{X}(X)=A_{X}{}^{\ast }A_{X}$ and the
reduced one-body density matrix $\rho _{X}(y,y^{\prime })$ is computed at
the central density $n_{X}.$ This $\rho _{X}(y,y^{\prime })$ is defined by
the formula (\ref{ro}) with $\psi _{n_{X}}=\psi _{X}$ which corresponds to
the homogeneous system $V_{n_{X}}$ of $N$ atoms with the density $n_{X}.$

Now we compute the $y$ derivatives of $\rho _{X\text{ }}.$ Setting $\partial
A_{X}(y)/\partial y=\partial A_{X}/\partial X$ and $\partial
^{2}A_{X}(y)/\partial y^{2}=\partial ^{2}A_{X}/\partial X^{2},$ the kinetic
energy $\Delta K_{X}$ (\ref{delKn}) of the subsystem with $\Delta N_{X}$
atoms and density $n_{X}$ obtains in the following form:%
\begin{equation}
\Delta K_{X}=-\frac{N\hbar ^{2}}{2m}\left( \frac{\Delta N_{X}}{Nf_{X}}%
\right) \int_{V_{X}}dydy^{\prime }\delta (y-y^{\prime })\left[ \left\vert
A_{X}\right\vert ^{2}\triangle _{y^{\prime }}\rho _{X}(y,y^{\prime })+\right.
\label{KX}
\end{equation}%
\begin{equation*}
\left. +\left( A_{X}^{\ast }\triangle _{X}A_{X}\right) \rho _{X}(y,y^{\prime
})+2A_{X}^{\ast }\mathbf{\nabla }_{X}A_{X}\cdot \mathbf{\nabla }_{y^{\prime
}}\rho _{X}(y,y^{\prime })\right] .
\end{equation*}

\subsection{Local equilibrium and the stationary states}

Now we can obtain the total energy $E$ (\ref{Esum}) as a functional of both
fast and slow modes. To this end, in $\Delta K_{X}$ (\ref{KX}) we return to
the expression of $\rho _{X}$ (\ref{ro}) in terms of the function $\psi
_{n_{X}}=\psi _{X}$. Adding $\Delta U_{X}$ (\ref{delVn}) to this equation
and setting $\Delta N_{X}=Nf_{X}\Delta V_{X}$, one obtains the total energy $%
E$ in the form of the following sum:%
\begin{eqnarray}
E\{\psi _{X},A_{X}\} &=&\sum_{X}\Delta
V_{X}\int_{V_{n}^{N}}d^{N}xA_{X}^{\ast }\psi _{X}^{\ast }\widehat{H}%
A_{X}\psi _{X},  \label{E} \\
\widehat{H} &=&\widehat{h}-\frac{i\hbar }{m}\mathbf{\nabla }_{X}\cdot
\sum_{i=1}^{N}\widehat{\mathbf{p}}_{i}-\frac{N\hbar ^{2}}{2m}\triangle _{X},
\label{HX}
\end{eqnarray}%
where $\widehat{h}$ is the short range Hamiltonian (\ref{hh}) and $\widehat{%
\mathbf{p}}_{i}=-i\hbar \mathbf{\nabla }_{i}$ is the momentum operator in $%
x_{i}$ which both act only on $\psi _{X}(x),$ whereas the operators $\mathbf{%
\nabla }_{X}$ and $\triangle _{X}$ act on $A_{X}=A(X).$ This $E$ is a
functional of both fast variables $\psi _{X}(x),\psi _{X}^{\ast }(x),$ and
slow variables $A(X),A^{\ast }(X),$ both normalized to unity:%
\begin{eqnarray}
\int_{V_{n}^{N}}d^{N}x\psi _{X}^{\ast }\psi _{X} &=&1,  \label{norm} \\
\sum_{X}\Delta V_{X}A_{X}^{\ast }A_{X} &=&1.  \notag
\end{eqnarray}

We are now ready to impose a local equilibrium which is determined by the
local density $n_{X}$. As the variations $\delta \psi _{X}$ and $\delta
A_{X} $ are very different in their temporal and spatial scales, the
variational equations must be applied separately to each of them. As now the
total wave function is $A_{X}\psi _{X}$, the variational equation for the
fast component at constant slow component $A$\ is
\begin{equation}
i\hbar A_{X}\partial \psi _{X}/\partial t=\frac{\delta E}{A_{X}^{\ast
}\delta \psi _{X}^{\ast }},  \label{VEpsi}
\end{equation}%
which gives the following Schr\"{o}dinger equation:%
\begin{equation}
i\hbar \partial \psi _{X}/\partial t=\left( \widehat{h}-\frac{i\hbar }{m}%
\mathbf{\nabla }_{X}\ln A_{X}\cdot \sum_{i=1}^{N}\widehat{\mathbf{p}}%
_{i}\right) \psi _{X}(x).  \label{SEpsi1}
\end{equation}%
By our assumption, locally the system is in a stationary states for which
the Schr\"{o}dinger equation (\ref{SEpsi1}) reduces to the eigen problem
\begin{equation}
\left( \widehat{h}-\frac{i\hbar }{m}\mathbf{\nabla }_{X}\ln A_{X}\cdot
\sum_{i=1}^{N}\widehat{\mathbf{p}}_{i}\right) \psi _{X}=N\varepsilon
_{X,A}\psi _{X}.  \label{eXA}
\end{equation}%
As $A_{X}$ is slow, the second term can be treated as a perturbation of the
Hamiltonian $\widehat{h}$. For this reason, the function $\psi _{X}$ will be
considered as the eigenfunction of the unperturbed operator $\widehat{h}$
with the eigenvalue $N\varepsilon _{X}$ which corresponds to the homogeneous
density $n_{X}$ as defined in eq.(\ref{epsX}). Then the total perturbed
eigenvalue $\varepsilon _{X,A}$ per atom is%
\begin{equation}
\varepsilon _{X,A}=\varepsilon _{X}-\frac{i\hbar }{m}\mathbf{\nabla }_{X}\ln
A_{X}\cdot \left\langle \mathbf{p}\right\rangle _{X},  \label{eXp}
\end{equation}%
and%
\begin{equation}
N\left\langle \mathbf{p}\right\rangle _{X}=\int_{V_{n}^{N}}d^{N}x\psi
_{X}^{\ast }\sum_{i=1}^{N}\widehat{\mathbf{p}}_{i}\psi _{X},  \label{p}
\end{equation}%
where $\left\langle \mathbf{p}\right\rangle _{X}$ is the corresponding
average momentum per particle (at the density $n_{X}$). If gas in a
subvolume $\Delta V_{X}$ is in the ground state then of course $\left\langle
\mathbf{p}\right\rangle _{X}=0,$ but in general the average momentum per
atom in $\Delta V_{X}$ is nonzero and depends on the density $%
n_{X}=N\left\vert A_{X}\right\vert ^{2}$, i.e., $\left\langle \mathbf{p}%
\right\rangle _{X}=\left\langle \mathbf{p}\right\rangle _{X}(\left\vert
A_{X}\right\vert ^{2}).$

So far we have assumed that the stationary state is a pure eigenstate, but
the stationary state can also be a superposition of more than one pure
states with an obvious weighted form of the average energy $\varepsilon _{X}$
per atom. For instance, a two state superposition is
\begin{equation}
\psi _{X}(x,t)=a_{1}\psi _{1,X}(x)e^{-iN\varepsilon _{1}t/\hbar }+a_{2}\psi
_{2,X}(x)e^{-iN\varepsilon _{2}t/\hbar },  \label{1+2}
\end{equation}%
where $\psi _{1,X}$ is the eigenstate with the per atom eigen value $%
\varepsilon _{X}=\varepsilon _{1}$, $\psi _{2,X}$ is the eigenstate with $%
\varepsilon _{X}=\varepsilon _{2},$ $t$ is the fast time, and $\left\vert
a_{1}\right\vert ^{2}+\left\vert a_{2}\right\vert ^{2}=1.$ In particular, $%
\psi _{1,X}$ can be a ground state with zero momentum and $\psi _{2,X}$ can
be a state with nonzero $\left\langle \mathbf{p}\right\rangle $. We see that
the two states have different exponential time dependence. However on
averaging over fast time, these exponentials play no role in the HA as they
disappear in the final hydrodynamic energy functional (\ref{EEsr}). Instead
of wave function of the form (\ref{1+2}) the expressions of the HA derived
below contain the effective wave function of the form
\begin{equation}
\psi _{X}=a_{1}\psi _{1,X}(x)+a_{2}\psi _{2,X}(x)  \label{1+2 ef}
\end{equation}%
and the average per atom energy $\varepsilon _{X}=\left\vert
a_{1}\right\vert ^{2}\varepsilon _{1}+\left\vert a_{2}\right\vert
^{2}\varepsilon _{2}.$ The case of a superposition of two states can be
directly generalized to the superposition of any number of eigenstates. For
simplicity, in what follows we will continue our consideration in terms
appropriate for a single pure inner state.

\subsection{The hydrodynamic equation}

Now we are ready to establish the equation describing the slow component $%
A_{X}$. Making use of eqs.(\ref{eXA}) and (\ref{eXp}) allows one to perform
averaging over the fast time $t$ and the $x$ integration in each $\Delta
V_{X}$ to eliminate the fast mode $\psi _{X}$. Then we change $\Delta V$ for
$dX$ and obtain the total energy $E_{sr}$ of the short range interaction as
a functional of $A^{\ast }$ and $A$ in the form of the\ following $X$
integral:%
\begin{equation}
E_{sr}\{A_{X}\}=N\int dX\left[ -\frac{\hbar ^{2}}{2m}A_{X}^{\ast }\triangle
_{X}A_{X}+\left\vert A_{X}\right\vert ^{2}e_{X}-\frac{i\hbar }{m}A_{X}^{\ast
}\mathbf{\nabla }_{X}A_{X}\cdot \left\langle \mathbf{p}\right\rangle _{X}%
\right] .  \label{EEsr}
\end{equation}%
If the external potential $U_{ext}$ changes rapidly, i.e., over the scale of
an order of the interparticle distance, then it has to be included in the
short range Hamiltonian operator $\widehat{h},$ otherwise it can be
incorporated along with the long range interaction potential $U_{lr}$ which
is assumed here. A long range potential changes over the large $X$ scale, it
interacts with the total number of particles $N\left\vert A_{X}\right\vert
^{2}dX$ in $\Delta V_{X}$ and its source in another subsystem $X^{\prime }$
is the total number of particles $N\left\vert A_{X^{\prime }}\right\vert
^{2}dX^{\prime }$ in $\Delta V_{X^{\prime }}.$ If both $U_{ext}$ and $U_{lr}$
are long range interactions, then the total energy due to this interactions
have the following form:%
\begin{eqnarray}
E_{lr}\{A_{X}\} &=&\frac{N(N-1)}{2}\int dX\int dX^{\prime }\left\vert
A_{X}\right\vert ^{2}\left\vert A_{X^{\prime }}\right\vert
^{2}(1+g_{2,XX^{\prime }})U_{lr}(X-X^{\prime })  \label{ELR} \\
&&+N\int dX\left\vert A_{X}\right\vert ^{2}U_{ext}(X-X^{\prime }),  \notag
\end{eqnarray}%
where $g_{2,XX^{\prime }}$ is the coarse-grained hydrodynamic pair
correlation obtained by averaging $G_{2}$ (\ref{G2}) over $\Delta V_{X}$ and
$\Delta V_{X^{\prime }},$ Appendix B. The total energy of the inhomogeneous
system is the sum $E=E_{sr}+E_{lr}.$ The dynamics of the slow modulation $%
A_{X}$ can be obtained from the following variational equation:%
\begin{equation}
i\hbar \sqrt{N}\partial A_{X}/\partial t=\frac{\delta E}{\sqrt{N}\delta
A_{X}^{\ast }}.  \label{VEA}
\end{equation}%
Performing this variation we have to remember that the local energy eigen
value $\varepsilon _{X}$ and average momentum $\left\langle \mathbf{p}%
\right\rangle _{X}$ depend on the local probability density $\left\vert
A_{X}\right\vert ^{2}.$ The coarse-grained hydrodynamic equation (\ref{VEA})
obtains in the form%
\begin{eqnarray}
i\hbar \frac{\partial A}{\partial t} &=&-\frac{\hbar ^{2}}{2m}\triangle
A+A\left( \varepsilon +\left\vert A\right\vert ^{2}\frac{\partial
\varepsilon }{\partial \left\vert A\right\vert ^{2}}\right) -\frac{i\hbar }{m%
}\mathbf{\nabla }A\cdot \left( \left\langle \mathbf{p}\right\rangle
+\left\vert A\right\vert ^{2}\frac{\partial \left\langle \mathbf{p}%
\right\rangle }{\partial \left\vert A\right\vert ^{2}}\right)  \label{SEA} \\
&&+AU_{ext}+(N-1)A\int_{V}dX^{\prime }\left\vert A_{X^{\prime }}\right\vert
^{2}(1+g_{2,XX^{\prime }})U_{lr}(X-X^{\prime }),  \notag
\end{eqnarray}%
where the arguments $X$ and $t$ of the functions $A$ and $U_{ext}$ are
omitted for brevity. Here $\varepsilon $ and $\left\langle \mathbf{p}%
\right\rangle $ are respectively the per atom eigenvalue of the operator $%
\widehat{h}$ (\ref{hh}) and the average particle momentum (\ref{p}) in a
homogeneous system of density $N\left\vert A\right\vert ^{2}$, which are
certain functions of $A^{\ast }A=\left\vert A\right\vert ^{2}.$ The second
and third terms in the rhs of the equation (\ref{SEA}) are the contributions
to the local chemical potential respectively from the unperturbed local
eigenvalue $\varepsilon $ and its perturbation: $\mu _{X}=\partial
(N_{X}\varepsilon _{X,A})/\partial N_{X}$ where $N_{X}=N\left\vert
A_{X}\right\vert ^{2}dX$ is the number of atoms in $\Delta V_{X}$ and $%
\varepsilon _{X,A}$ is the total eigenvalue (\ref{eXp}). This derivative is
equal to%
\begin{equation}
\mu _{X,A}=\partial (\left\vert A_{X}\right\vert ^{2}\varepsilon
_{X,A})/\partial \left\vert A_{X}\right\vert ^{2},  \label{muX}
\end{equation}%
which results in the terms in question. It is this $\mu _{X}$ that
universally appears in the HE and, in this sense, plays the role similar\ to
the pressure gradient in the classical HEs. The specific form of its $A$
dependence determines the nonlinearity in $A$ and depends on the specific
system via the specific density dependence of the local eigenvalues of the
stationary Schr\"{o}dinger equation (\ref{eXA}).

Equation (\ref{SEA}) shows that the flows of the fast and slow modes
interact via the product $\propto \mathbf{\nabla }A\cdot \left\langle
\mathbf{p}\right\rangle $. If $e_{X}$ is the eigenvalue of an excited state
with a nonzero momentum, then this term can play its role. The GPE obtains
from (\ref{SEA}) for $\left\langle \mathbf{p}\right\rangle =0$ in the weak
interaction limit: both in 3D \cite{LHY} and 1D \cite{LL,L} the first term
in the energy expansion is proportional to the density $n\propto \left\vert
A\right\vert ^{2}$ so that $\varepsilon \left\vert A\right\vert ^{2}\propto
\left\vert A\right\vert ^{4},$ which results in the GPE with the qubic
nonlinearity. If $\varepsilon $ is the ground state eigenvalue for the
Lieb-Lineger gas, then (\ref{SEA}) gives the equation considered in \cite%
{Olshanii2001,Santos2002,Kim2003,PawlPRL,Pawl SciPost}. In particular, this
approach reduces to the eGPE for a low density and to the fifth-power
nonlinearity proposed in \cite{Kolom} in the strong coupling regime.
However, the equation (\ref{SEA}) is not restricted to a specific model and
dimension and is a general HE for a cold Bose gas.

The equation (\ref{SEA}) can be presented in the classical hydrodynamic
form. Setting $A=a\exp (i\theta /\hbar )$ with the real amplitude $a$ and
phase $\theta $, we multiply (\ref{SEA}) by $a\exp (-i\theta /\hbar )$ and
separate the imaginary and real parts. The imaginary part can be reduced to
the form%
\begin{equation}
\frac{\partial a^{2}}{\partial t}+\mathbf{\nabla }\left[ a^{2}(\mathbf{v}%
+\left\langle \mathbf{p}\right\rangle /m)\right] =0,  \label{Cont}
\end{equation}%
where $\mathbf{v}=\mathbf{\nabla }\theta /m$. This equation is interpreted
as that of continuity and $\mathbf{v}$ as velocity of the slow mode. The
real part can be interpreted as the Hamilton-Jacobi equation with the
quantum corrections \cite{Landau}. Applying operator $\mathbf{\nabla }$ to
this equation, one obtains the analog of the standard hydrodynamic
Navier-Stokes equation for an inviscid fluid:
\begin{equation}
\frac{\partial \mathbf{v}}{\partial t}+(\mathbf{v\nabla })\mathbf{v=-}\frac{1%
}{m}\mathbf{\nabla }\left[ \mu _{tot}+\mathbf{v\cdot }\left( \left\langle
\mathbf{p}\right\rangle +a^{2}\frac{\partial \left\langle \mathbf{p}%
\right\rangle }{\partial a^{2}}\right) \right] ,  \label{Euler}
\end{equation}%
where
\begin{equation}
\mu _{tot}=\varepsilon +\left\vert A\right\vert ^{2}\frac{\partial
\varepsilon }{\partial \left\vert A\right\vert ^{2}}-\frac{\hbar
^{2}\triangle a}{2ma}+V_{ext}+(N-1)\int_{V}dX^{\prime }\left\vert
A_{X^{\prime }}\right\vert ^{2}U_{lr}(X-X^{\prime })(1+g_{2,XX^{\prime }})
\label{mu}
\end{equation}%
and $\left\langle \mathbf{p}\right\rangle $ is the average momentum per
atom. We see that the rhs has the form (pressure gradient/$N)=\mathbf{\nabla
}(\mu _{X,A}+\mu _{q}+\mu _{lr})/m$ where $\mu _{q}\sim \hbar ^{2}/m$ is the
quantum pressure term and $\mu _{lr}$ is the contribution to the local
chemical potential due to the long range interaction. Note that the relation
between Schr\"{o}dinger's equation and hydrodynamics has been revealed by
Madelung long ago \cite{Madelung} and since then has been extensively used
as the motivation for the HA and LDA. The hydrodynamic Madelung
representation is derived from Schr\"{o}dinger's equation for the
one-particle wave function \cite{Madelung,Landau}. In contrast, we arrived
at the above HEs (\ref{Cont}) and (\ref{Euler}) in the reverse procedure,
i.e., first deriving the HE (\ref{SEA}) from the many-particle Schr\"{o}%
dinger's equation (\ref{eXA}) and only then applying Madelung's ansatz to
the slow mode $A$ which is indeed a one-particle quantity$.$

\section{The total effective wave function.}

\subsection{Introduction of the common coordinate}

The general energy functional (\ref{E}) can be presented in the following
suggesting form:%
\begin{eqnarray}
E_{sr}\{\Psi \} &=&\int_{V}dX\int_{V_{X}^{N}}d^{N}x\Psi _{X}^{\ast }(x)%
\widehat{H}\Psi _{X}(x),  \label{EE} \\
\Psi _{X}(x) &=&A_{X}\psi _{X}(x).  \label{PSI}
\end{eqnarray}%
Here $\Psi _{X}(x)$ is the total effective inhomogeneous wave function and $%
E_{sr}\{\Psi \}$ can be interpreted as the expectation value of the
Hamiltonian $\widehat{H}$ (\ref{HX}) in the state $\Psi _{X}(x).$ The
function $\psi _{X}$ is of the type (\ref{1+2 ef}) without fast $t$
exponentials. The effective wave function depends on the $Nd$ dimensional
vector $x=(x_{1},x_{2},...,x_{N})$ and the coarse-grained coordinate $X.$
The $d$ dimensional components $x_{i}$ of the argument $x$\ in $\Psi _{X}(x)$
are defined in the auxiliary volume $V_{n_{X}}$, Fig.1, which we denote
simply as $V_{X},$ and $X$ is a $d$ dimensional vector defined in the actual
system's volume $V.$ The dependence of the effective wave function $\Psi
_{X}(x)=A_{X}\psi _{X}(x)$ on the coarse-grained coordinate $X$ through $%
A_{X}$ is explicit, and is in general functional one via dependence of $\psi
_{X}(x)$ on the local density $n_{X}$. Thus, each local subvolume $\Delta
V_{X}$ is represented by a homogeneous system $V_{X}$ of $N$ atoms at
density $n_{X},$ the number of atoms in $\Delta V_{X}$ is set by the factors
$A_{X}^{\ast }$ and $A_{X},$ so that the operator $\widehat{H}$ acts on the
vector $X\oplus x\in \Delta V_{X}\otimes V_{X}^{N}.$\ The total space in
which our inhomogeneous gas is described is $\oplus _{X}(\Delta V_{X}\otimes
V_{X}^{N}).$ It is appropriate to associate $X$ and $\Delta V_{X}$ with the
macrospace and $x$ and $V_{X}$ with the inner space: in the inner space the
system is in the stationary quantum state $\psi _{X}(x)$ and in the
macrospace space the system is in the state $A_{X}.$

The coordinates of the inner vectors $x$ in different subsystems $\Delta
V_{X}$ are independent as they are defined in different domains $V_{X}$ and
so far there was no need in their "synchronization". However, after the
inner space has been used and the HE (\ref{SEA}) already obtained, there is
no need to resort to the auxiliary domains $V_{X}$, but instead there is a
need of description in terms of the atoms' coordinates in the actual volume $%
V.$ Then it is convenient and, for some reason even necessary, to use the
common coordinate system in which vector $\widetilde{x}=(\widetilde{x}%
_{1},...,\widetilde{x}_{N})$ has components that are smoothly interpolating
those of the local inner vectors $x^{(X)}$ in different subsystems $\Delta
V_{X}$ through the entire system. For instance, the phase of the total wave
function depends on the atoms' positions in the actual space $V$ and thus is
a function of such $\widetilde{x}$ with the components $\widetilde{x}_{i}$
from $V.$ To introduce $\widetilde{x}$, the inner coordinates $x$ in
different subsystems $\Delta V_{X}$ can be "synchronized" by choosing the
common coordinate onset in all $V_{X}$ as sketched in Fig.1. Let $%
x^{(X)}=(x_{1}^{(X)},x_{2}^{(X)},...,x_{N}^{(X)})$ be the vector of a point
related to $\Delta V_{X}$ and defined in its $V_{X}.$ If the onset is
common, then at the boundary $B_{12}$ between two adjacent $\Delta V_{X_{1}}$
and $\Delta V_{X_{2}}$, the components of $%
x^{(2)}=(x_{1}^{(2)},x_{2}^{(2)},...,x_{N}^{(2)})$ with $x_{i}^{(2)}\in
\Delta V_{X_{2}}$ are equal to the components of $%
x^{(1)}=(x_{1}^{(1)},x_{2}^{(1)},...,x_{N}^{(1)})$ with $x_{i}^{(1)}\in $ $%
\Delta V_{X_{1}};$ similar matching takes place at all boundaries between
neighboring $\Delta V$'s. Thanks to this pairwise adjustment, combining
coordinates in different $\Delta V_{X}$ in the common reference frame one
obtains the vector $\widetilde{x}=(\widetilde{x}_{1},\widetilde{x}_{2},...,%
\widetilde{x}_{N})$ that continuously interpolates between the local
coordinates through the entire system. The vector $\widetilde{x}$ does not
need the superscript $X$ and is such that i) its components $\widetilde{x}%
_{i}$ are continuous and defined in the total system's volume $V,$ ii) its
components $\widetilde{x}_{i}$ in $\Delta V_{X}$ coincide with the local
components $x_{i}^{(X)}$ of this point , i.e., $\widetilde{x}%
_{i}=x_{i}^{(X)} $ if $\widetilde{x}_{i}\in \Delta V_{X}$. As the densities
at both sides of the boundaries $B_{XX^{\prime }}$ are equal, $\psi _{X}(%
\widetilde{x})$ is a continuous function of the variable $\widetilde{x}$. In
particular, the phase of $\psi _{X}(\widetilde{x})$, $arg(\psi _{X}),$ is a
function of the coordinates of all atoms in the actual volume $V.$ In what
follows we omit tilde and use notation $x$ for the common coordinate $%
\widetilde{x}.$

\subsection{ Two forms of states with a nonzero momentum}

As usual, in the HA the system is specified in its inner space, and here we
present two important examples of inner states (\ref{PSI}) which we are
going to implement below. For clarity, the $d$ dimensional vectors such as
components $x_{i}$ of $x$, as well as $y$ and $X$ will be presented in bold
face.

First, we want to set gas in motion as a whole. Let the local ground state
wave function of the initial state be $\Psi _{X,0}=A_{0}(\mathbf{X,}t)\psi
_{X,0}(x).$ In this state, the energy density is $N\left\vert
A_{X}\right\vert ^{2}\varepsilon _{X}$ where $\varepsilon _{X}$ is given in (%
\ref{epsX}), momentum density is $0,$ and $A_{0}$ is governed by the
equation (\ref{SEA}) with $\left\langle \mathbf{p}\right\rangle =0.$ If this
gas is set in motion as a whole so that all atoms have the same momentum $%
\mathbf{p},$ then its new inner state $\psi _{X,p}$ and hence the total $%
\Psi _{X,p}$ acquires the factor $\exp \sum_{i}^{N}i(\mathbf{px}_{i})/\hbar $
where $(\mathbf{px}_{i})$ is the scalar product of two $d$ dimensional
vectors. Then the total wave function of the gas moving as a whole is%
\begin{eqnarray}
\Psi _{X,p} &=&A_{X}\psi _{X,p}(x)  \label{Psi1} \\
&=&A_{X}\psi _{X,0}(x)\exp \sum_{i=1}^{N}i(\mathbf{px}_{i})/\hbar .  \notag
\end{eqnarray}%
This $\Psi _{X,p}$ is the local state with the energy density $N\left\vert
A_{X}\right\vert ^{2}(\varepsilon _{X,0}+$ $p^{2}/2m)-N\frac{i\hbar }{m}%
A_{X}^{\ast }(\mathbf{p\cdot \nabla }_{X})A_{X}$ where $\varepsilon _{X,0}$
is the eigenvalue for the eigenfunction $\psi _{X,0}$, it has the inner
momentum $N\mathbf{p}$ with density $N\mathbf{p}\left\vert A_{X}\right\vert
^{2};$ the macro state $A$ is now the solution of the equation (\ref{SEA})
with $\left\langle \mathbf{p}\right\rangle =\mathbf{p}.$ Thus, in the state (%
\ref{Psi1}), the momentum density is changing along with the atom density $%
\propto \left\vert A_{X}\right\vert ^{2}$, but the individual momenta $%
\mathbf{p}=const.$ As the ground state function is real and its phase is
zero, the fast phase at some point $\mathbf{y}$ in the gas volume $V$ is
equal to $\theta _{fm}(\mathbf{x}_{k}=\mathbf{y})=\arg [\exp iN(\mathbf{py}%
)/\hbar ]=N(\mathbf{py})/\hbar $ and the total phase $\theta _{tot}(\mathbf{y%
})=\theta _{sm}+\theta _{fm}$ where $\theta _{sm}(\mathbf{y})=\arg [A(%
\mathbf{y})].$ It can be noticed that in this state the fast phase is
additive so that it is possible and natural to assign an individual fast
phase $(\mathbf{px}_{i})/\hbar $ to each atom. Then the total phase of a
single atom is $\theta _{1}(\mathbf{y})=\theta _{sm}(\mathbf{y})+(\mathbf{py}%
)/\hbar $. The related problem is that in a ring geometry stretched, say,
along $y$, the periodic boundary condition \ implies that the total phase
change $\theta _{tot}(L)$ along the system length $L$ must be a multiple of $%
2\pi .$ However, if $\Delta \theta _{sm}=\theta _{sm}(L)\neq 0,$ which is
the case of gray solitons, this phase $\theta _{tot}(L)=$ $\Delta \theta
_{sm}+NpL/\hbar $ is different and the excess phase $\func{mod}(\theta
_{tot},2\pi )$ must be compensated. Our result shows that the compensating
source can be a very slow flow of the gas as a whole with the velocity $%
v^{\prime }$ chosen as to cancel the phase, i.e., to make $\func{mod}[\theta
_{tot}(L)+$ $Nmv^{\prime }L/\hbar ,2\pi ]=$ $0$. Due to additivity of the
fast phase, this periodic boundary condition reduces to the condition $\func{%
mod}[\theta _{1}(L)+$ $v^{\prime }mL/\hbar ,2\pi ]=$ $0$ on the phase $%
\theta _{1}(y)$ of single atom, which is equivalent to
\begin{equation}
\Delta \theta _{sm}+\func{mod}[(p+mv^{\prime })L/\hbar ,2\pi ]=0.
\label{mod1}
\end{equation}%
While the additional kinetic energy $Nmv^{\prime 2}/2\sim 1/N$ is
negligible, the total slow momentum per atom $p_{sm}$, up to unimportant
constant, changes to
\begin{equation}
p_{sm}^{\prime }=p_{sm}-\Delta \theta _{sm}\hbar /L.  \label{psm}
\end{equation}%
\emph{\ }This expression for the new momentum $p_{sm}^{\prime }$ is in line
with the so-called momentum renormalization which has always been assumed in
the soliton theory \cite{Fad,Pit,Kivshar}. The advantage of dealing only
with the single atom phase $\theta _{1}$ makes it possible the description
in terms of a one-particle wave function. It is shown below that, both in
our HA and in the standard HA, a gas moving as a whole can indeed be
described by the same equation for a one-particle wave function.

Second, we want to describe a state with average nonzero momenta of some $%
\alpha N$ out of $N$ atoms. In this state, the total wave function is
similar to (\ref{1+2 ef}), i.e., it is a superposition of the local ground
state $A_{X}\psi _{X,0}(x)$ and state with certain momentum :%
\begin{eqnarray}
\Psi _{X,p} &=&A_{X}\psi _{X,p}(x)  \label{Psi2} \\
&=&A_{X}\psi _{X,0}(x)\left[ a_{0}+a_{p}\exp \sum_{i=1}^{N}i(\mathbf{px}%
_{i})/\hbar \right] ,  \notag
\end{eqnarray}%
where $(a_{p}/a_{0})^{2}=\alpha .$ If the ground state as a whole is set in
motion, then $a_{0}=0,a_{p}=\alpha =1,$ and the wave function (\ref{Psi2})
reduces to the wave function (\ref{Psi1}). The state (\ref{Psi2}) has the
energy density $N\left\vert A_{X}\right\vert ^{2}(\varepsilon _{X,0}+$ $%
\alpha p^{2}/2m)-N\frac{i\hbar }{m}A_{X}^{\ast }(\alpha \mathbf{p\cdot
\nabla }_{X})A_{X}$ where $\varepsilon _{X,0}$ is the eigenvalue for the
eigenfunction $\psi _{X,0},$ the inner momentum density is $N\alpha \mathbf{p%
}\left\vert A_{X}\right\vert ^{2}.$ In each $\Delta V$ with its $\left\vert
A\right\vert ^{2},$ the average ratio (number of atoms with momentum $%
\mathbf{p}$)/(number of atoms with zero momentum) $=$ $\alpha $ and remains
fixed. Now the fast phase at a point $\mathbf{y}$ in $V$ is $\theta _{fm}(%
\mathbf{y})=\arg [\psi _{X,p}(\mathbf{x}_{i}=\mathbf{y})]:$
\begin{equation}
\theta _{fm}(\mathbf{y})=\arctan \left[ \frac{a_{p}\sin \left[ N(\mathbf{py}%
)/\hbar \right] }{a_{0}+a_{p}\cos \left[ N(\mathbf{py})/\hbar \right] }%
\right] .  \label{Teta fm}
\end{equation}%
This phase cannot be separated into individual phases of single atoms so
that, in contrast to the above case of gas motion as a whole, the
one-particle description is impossible. The counterpart of equation (\ref%
{mod1}) for the velocity $v^{\prime }$ of the phase compensating flow in a
one-dimensional ring geometry has the form
\begin{equation}
\func{mod}[\Delta \theta _{sm}+\theta _{fm}(L)+Nmv^{\prime }L/\hbar ,2\pi
]=0,  \label{mod2}
\end{equation}%
where $\theta _{fm}$ is given in (\ref{Teta fm}). In spite of the difference
with the case of the wave function (\ref{Psi1}), the renormalized slow mode
momentum is the same, eq.(\ref{psm}). The states with the wave function (\ref%
{Psi2}) will be addressed in Sec.IV\textbf{. }

The two states (\ref{Psi1}) and (\ref{Psi2}) are useful for our presentation
as their many-particle phases are known. At the same time, the phase of a
single excited eigenstate of the operator $\widehat{h}$ (\ref{hh}) with
nonzero momentum is usually not known. Such a state is employed in Sec.VI
where the momentum and energy eigenvalues will suffice for our purpose. The
peculiarity of such states is that the number of excited atoms remains
constant in all $\Delta V$, but their momenta can depend on the density via $%
\left\vert A\right\vert ^{2}.$

\subsection{General properties of the general HA and its relation to the
standard HA}

As a general HA, the presented description of a cold Bose gas comprises two
modes. As a result, not only the energy (\ref{EEsr}), but also the total
momentum $\mathbf{P}$ and total phase $\theta $ consist of both slow and
fast contributions:%
\begin{eqnarray}
\mathbf{P} &=&\mathbf{P}_{sm}+N\left\langle \mathbf{p}\right\rangle ,
\label{P and teta} \\
\theta _{tot} &=&\theta _{sm}+\theta _{fm},  \notag
\end{eqnarray}%
where $\mathbf{P}_{sm}=-i\hbar mN\int dV(A^{\ast }\mathbf{\nabla }A-A\mathbf{%
\nabla }A^{\ast })$ is the slow mode momentum and $N\int dV\left\langle
\mathbf{p}\right\rangle _{X}/V$ is the fast mode momentum, the phases are
defined at points $\mathbf{y}$ in the actual volume $V$, $\theta _{sm}(%
\mathbf{y})=\arg A(\mathbf{y})$ and $\theta _{fm}(\mathbf{y})=\arg [\psi
_{X}(\mathbf{x}_{i}=\mathbf{y})]$. We see that our HA which explicitly
describes an inhomogeneous system in terms of both microscopic inner
many-particle fast mode $\psi (x)$ and the slow macroscopic hydrodynamic
mode $A(X)$ is substantially different from the standard HA which describes
the system solely in terms of the slow one-particle mode. We will now show
that nevertheless, at least in a one-dimensional geometry, both descriptions
are consistent with one another as long as the gas moves as a whole and $%
\left\langle p\right\rangle $ is a constant momentum $p$ of every atom,
which is independent of the density.

To address this and our next problems we introduce the renormalized time $%
\tau $, coordinate $y,$ momentum $p,$ and new function $\phi (\tau ,y):$ $%
t=\tau m/(\pi ^{2}n_{0}^{2}\hbar ),$ $X=y/(\pi n_{0}),$ $x_{k}=y_{k}/(\pi
n_{0})$ where $y$ is the renormalized one-dimensional coordinate from $V$, $%
y_{i}$ is the renormalized coordinate of $i$ th atom, $p=\left\langle
p\right\rangle /(\pi n_{0}\hbar ),$ $\phi =A\sqrt{N/n_{0}},$ where $%
n_{0}=N/L $ is the unperturbed density and $L$ is system's length. In these
variables, the energy is in units $\hbar ^{2}\pi ^{2}n_{0}^{2}/m$ and $L=\pi
N$; the atom momentum, related velocity, which is now in units of the sound
velocity $v_{S}=\pi n_{0}\hbar /m,$ and wave number $\left\langle
p\right\rangle /\hbar $ in units $\pi n_{0}$ are equal numbers (which one
should keep in mind). For a constant $p,$ the general form of equation (\ref%
{SEA}) in one-dimension is%
\begin{equation}
i\partial _{\tau }\phi =-\frac{1}{2}\phi ^{\prime \prime }+\mu (|\phi
|^{2})\phi -ip\phi ^{\prime },  \label{HA}
\end{equation}%
where prime of $\phi $ stands for the $y$ derivative and $\mu (\left\vert
\phi \right\vert ^{2})=\varepsilon +\left\vert \phi \right\vert ^{2}\partial
\varepsilon /\partial \left\vert \phi \right\vert ^{2}$ is the unperturbed
local chemical potential. As we showed above, if $\psi _{g}$ is the initial
local inner ground state of the gas, but the gas is set in motion as a whole
and all atoms have the same momenta $p,$ then the $\psi _{g}$ acquires the
factor $\exp (\sum_{k}^{N}ipy_{k})$, the inhomogeneous system is described
by the total function of the form (\ref{Psi1}), i.e., $\Psi =\phi (y,\tau
)\psi _{g}\exp (\sum_{i}^{N}ipy_{i}),$ all atoms have similar fast phase $py$%
, and the slow mode $\phi $ is the solution of eq.(\ref{HA}). At the same
time, the corresponding equation of the standard HA is similar to (\ref{HA}%
), but without $p$ term:%
\begin{equation}
i\partial _{\tau }\phi =-\frac{1}{2}\phi ^{\prime \prime }+\mu (|\phi
|^{2})\phi .  \label{LDA}
\end{equation}%
However, now both slow and fast modes appear as the solution of the standard
equation (SE) (\ref{LDA}) which describes the one-atom state and therefore
must have the form $\phi _{SE}=\phi _{sm}(y,t)\exp i(py-p^{2}t/2)$ where $%
\phi _{sm}(y,t)$ is the slow mode and $p$ is an arbitrary momentum.
Substituting this $\phi _{SE}$ in the equation (\ref{LDA}) we find that the $%
\phi _{sm}$ satisfies exactly the equation (\ref{HA}) of our HA. Thus, the
slow component of $\phi _{SE}$ and $\phi $ satisfy the same equation (\ref%
{HA}), the total spatial phase of the function $\phi _{SE}$ and that of an
individual atom in $\psi $ are similar and equal to $\arg (\phi )+py$. The
difference is that both momentum $p$ and coordinate $y$ in the standard HA
and our HA have different origin. In the standard HA, $py$ is the common
linear phase of all atoms whereas in our HA, $py_{i}$ is the fast phase of $%
i $ th atom; the momentum $p$ in the standard HA is the parameter of the
solution $\phi _{SE}$ which is interpreted as the wave number of the
background oscillation \cite{Kivshar}, whereas in our HA, $p$ is the local
momentum eigenvalue per atom. Thus, the SE (\ref{LDA}) for the supposedly
macroscopic variable $\phi _{SE}$ describes both fast oscillations and
actual slow variable $\phi _{sm},$ the last alone being governed by our
equation (\ref{HA}). Thanks to this fact both the LDA and our HA are
equivalent in the case when the gas is moving as a whole. However, in the
state with the wave function (\ref{Psi2}) when certain gas fraction is in
the ground state and the total momentum is that of the local excitations,
and/or when this momentum depends on the local density, then such a gas can
be described only by our HA as the standard HA is not applicable in this
case. Below we illustrate possible effects in such systems by two examples.

\section{Can a periodic soliton train propagate?}

The GPE has both single soliton solution and the so-called periodic soliton
train solution \cite{Gagnon,Carr1}. Here we show that in contrast to the
former solution whose phase at the periphery can be of the form $v_{0}y$
with arbitrary $v_{0}$ independent of the soliton velocity $v$, the later is
always of the form of a standing wave
\begin{equation}
\phi _{train}=\phi (y-vt)\exp i(vy-v^{2}t/2).  \label{Fi train}
\end{equation}%
Here $\phi (y-vt)$ is the slow mode which describes the periodic train
moving with velocity $v$, and the exponential $\exp (ivy)$ describes the
total gas set in motion as a whole with the same velocity $v.$ The reason
why we call this a standing wave is that the train's velocity with respect
to the medium (the gas) is zero, i.e., the train does not propagate with
respect to the medium. We show that the above $\phi _{train}$ is a general
solution of the time dependent GPE.

We are interested in the solution of eq.(\ref{LDA}) in the form $\phi
_{train}=\phi (y,t)\exp i(v_{0}y-v_{0}^{2}t/2)$ in which $\phi $ is a
stationary solution $\phi =\phi (y-vt)=R\exp i\theta _{sm}$ where $R$ is the
real amplitude and $\theta _{sm}$ is real phase. As shown above, if the gas
is moving as a whole with velocity $v_{0}$ then the slow mode $\phi $ is the
solution of the equation (\ref{HA}) with $p=v_{0}$. The imaginary part of (%
\ref{HA}) is the phase equation which does not depend on the potential $\mu $
and has the form%
\begin{equation}
(\theta _{sm}^{\prime }+v_{0}-v)R^{2}=C,  \label{PhE}
\end{equation}%
where $C$ is an integration constant. If the solution is solitary, then, at
the periphery, $R$ takes a constant value $R_{\infty },$ $\theta
_{sm}^{\prime }=0,$ and one has $\theta _{sm}(y)=(v_{0}-v)\int_{-\infty
}^{y}dy^{\prime }[R_{\infty }^{2}/R^{2}(y^{\prime })-1].$ Adding the fast
phase $v_{0}y$ we obtain $\theta _{tot}=\theta _{sm}(y)+v_{0}y,$ which
describes the medium moving with the velocity $v_{0}$ and the soliton
superimposed upon it. As $v$ in $R=R(y-vt)$ is arbitrary, the soliton is
propagating in the medium with the relative velocity $v-v_{0}$ restricted
only by the speed of sound.

In the case of a periodic solution, the phase acquires a term linear in $y.$
For periodic $R$ eq. (\ref{PhE}) is solved by
\begin{equation}
\theta _{sm}(y)=\int_{0}^{y}dy^{\prime }C/R^{2}(y^{\prime })+(v-v_{0})y.
\label{Teta}
\end{equation}%
It has been shown that the first integral is expressed in terms of elliptic
functions and does not have terms linear in $y$ \cite{Gagnon,Carr1}$.$ The
total phase is then $\theta _{tot}=\theta _{sm}+v_{0}y$ and has the linear
term $vy$ which indicates that the resulting velocity of the medium is $v$
in accord with (\ref{Fi train}). Thus, the periodic train is moving with the
velocity $v$ exactly equal to that of the medium and therefore is a standing
wave. This means that the periodic soliton train can be set in motion with
velocity $v_{0}$ only along with the whole medium moving with the same
velocity and momentum $Nmv_{0}.$ The wave function of the state considered
above is of the form (\ref{Psi1}), it allows for the one-particle
description and can be considered both in the standard and our HA.

To make the train a propagating wave, its velocity must be made different
from the velocity of the ground state atoms. This is possible if the medium
has $\alpha N<N$ local excitations $v_{0}$ with small $\alpha $ so that the
total momentum $N\left\langle \mathbf{p}\right\rangle =\alpha Nv_{0}<Nv_{0}.$
In this case, the wave function is of the form (\ref{Psi2}), the slow mode
is governed by equation (\ref{HA}) with $p=\alpha v_{0}$ and the slow phase
has the form (\ref{Teta}) in which $v_{0}$ is replaced with $\alpha v_{0}.$
But now in average $N(1-\alpha )$ atoms remain in the ground state fraction
with zero momenta and only small fraction of $\alpha N$ atoms has average
momenta. The total wave function of the fraction of $N(1-\alpha )$ atoms
with zero momentum has the form $a_{0}R(y-vt)\exp (i\theta _{sm})\psi _{g}$
where $\theta _{sm}$ has the linear term $(v-\alpha v_{0})y.$ We see that
the train velocity $v$ differs from that of the medium: its major fraction
which in average consists of $N(1-\alpha )$ atoms is moving with the common
velocity $v-\alpha v_{0},$ hence the train drifts with respect to the medium
with the velocity $\alpha v_{0}.$ If $\alpha =1$ then $\theta _{fm}=vy$, the
velocities of the train and medium are both $v$, and the train is a standing
wave.\ Thus, the local excitation momenta can push the train with respect to
the bulk of atoms at rest thereby making the soliton train a propagating
wave. The state considered cannot be described in terms of the one-particle
HA.

\section{The effect of the density dependent local momenta on a soliton}

A stationary soliton with velocity $v$ in a gas with a constant momentum $%
\delta $ per atom (both due to all atoms in the same motion and only to some
atoms with nonzero momenta) has the form $\phi (y-vt|v,\delta )$ which
depends on the parameters $v$ and $\delta .$ This solution can be obtained
from that in a gas at rest, $\phi (y-vt|v),$ by a simple shift of the
parameter $v$ by $\delta ,$ i.e., $\phi (y-vt|v,\delta )=\phi (y-vt|v-\delta
)$. This is general rule that follows from this observation: both in the
imaginary part [equation (\ref{Teta})] and in the real part of eq.(\ref{HA}%
), the parameter $v$ appears in the combination $v-\delta .$ But we now want
to see how a soliton solution can be modified in a gas with local momenta
that depend on the local gas density. To this end, as a simple application
of the obtained equation (\ref{SEA}), we consider a cold one-dimensional
Lieb-Liniger Bose gas in the strong repulsion Tonks-Girardeau regime \cite%
{LL,L}. The energy structure of a Lieb-Liniger gas in this regime is known
to be Fermilike with the ground state in which all the energy levels up to
certain maximal value are occupied. We assume that a small fraction of $%
\alpha N$ upper levels are excited and bear momenta $p.$ In principle, this $%
p$ can depend on the local density in itself. To support this possibility,
we suggest the following reasoning. The total momentum of a Lieb-Liniger gas
is a multiple of $2\pi /L$ \cite{LL}$.$ As the inner state $\psi _{X}$ of
each subvolume $\Delta V_{X}$ is related to a homogeneous system of $N$
atoms and length $L_{X}=N/n_{X},$ the possible momenta in a subvolume $%
\Delta V_{X}$ are multiples of $2\pi n_{X}/N$ which is proportional to the
density $n_{X}.$ If atoms' momenta $p$ are adiabatically following the
momentum levels that are slowly changing in space along with density, then $%
p(\left\vert A\right\vert ^{2})\propto \left\vert A\right\vert ^{2}.$ This
suggests a model of an inhomogeneous system, in which both energy and
momentum of individual atoms are changing with the density.

Consider the eigenstate with $\alpha N$ exited atomic levels. If $p$ is the
excitation momentum at the nonperturbed density $n_{0}$, then the excitation
momentum at density $n$ is $(n/n_{0})p.$ The ground state energy per atom is
$e_{0}=(n/n_{0})^{2}/6$ \cite{LL,L,Excitation energy}. As the density is $%
n=N\left\vert A\right\vert ^{2}=n_{0}|\phi |^{2},$ one has $e_{0}=\left\vert
\phi \right\vert ^{4}/6$ and $e_{0}+\left\vert \phi \right\vert ^{2}\partial
e_{0}/\partial \left\vert \phi \right\vert ^{2}=$ $\left\vert \phi
\right\vert ^{4}/2,$ $\left\langle p\right\rangle (n)=\delta \left\vert \phi
\right\vert ^{2}$ and $\left\langle p\right\rangle +\left\vert \phi
\right\vert ^{2}\partial \left\langle p\right\rangle /\partial \left\vert
\phi \right\vert ^{2}=2\delta \left\vert \phi \right\vert ^{2}$ where $%
\delta =\alpha p$. Making use of these functions in the energy (\ref{EEsr}),
one gets%
\begin{equation}
E=\frac{1}{\pi }\int_{0}^{L}dy\left( -\frac{1}{2}\phi ^{\ast }\phi ^{\prime
\prime }+\frac{1}{6}\left\vert \phi \right\vert ^{6}-i\delta \phi ^{\ast
}\phi ^{\prime }\left\vert \phi \right\vert ^{2}\right) .  \label{Esol}
\end{equation}%
Substituting $\phi (y-v\tau )\exp (-i\lambda \tau )$ in the HE (\ref{HA})
gives the equation for $\phi :$
\begin{equation}
iv\phi ^{\prime }-\frac{1}{2}\phi ^{\prime \prime }+\frac{1}{2}|\phi
|^{4}\phi -2i\delta \phi ^{\prime }\left\vert \phi \right\vert ^{2}-\lambda
\phi =0.  \label{EQ}
\end{equation}%
The function $\phi $ has the asymptotics $\phi (y\rightarrow \pm \propto )=1$
$[A(x\rightarrow \pm \propto )=1/\sqrt{L}]$ so that $\int dy\left\vert \phi
\right\vert ^{2}=\pi N$ . We are looking for the stationary solution of eq.(%
\ref{EQ}) in the form $\phi (\tau ,y)=R(y-v\tau )e^{i\theta (y-v\tau )}$
where $R$ and $\theta $ are real functions, and $v$ is the wave velocity in
units $v_{s}=\pi n_{0}\hbar /m$ of sound velocity for $\delta =0$. The
soliton profile $\left\vert \phi \right\vert ^{2}$ is found to be
\begin{equation}
\left\vert \phi \right\vert ^{2}=1-\frac{3B}{2+3v\delta +D\cosh [2\sqrt{B}%
(y-v\tau )]},  \label{amplitude}
\end{equation}%
where $B=1-(v-\delta )^{2}+2v\delta ,$ $D=\sqrt{(1+3v^{2})(1+3\delta ^{2})}.$
The phase change $\Delta \theta _{sm}$ along the soliton due to the slow
mode is%
\begin{equation}
\Delta \theta _{sm}=-sign(v-\delta )\cos ^{-1}\left( \frac{3(v-\delta
)^{2}-1-3v\delta }{D}\right) +\sqrt{3}\delta \ln \left( \frac{\sqrt{3B}%
+2+3v\delta }{D}\right) .  \label{slow teta}
\end{equation}%
The condition $B\geq 0$ yields the following anisotropic restriction on the
soliton velocity which is different in the directions along and opposite to $%
\delta $ : $2\delta -\sqrt{1+3\delta ^{2}}<v<2\delta +\sqrt{1+3\delta ^{2}}$%
. It is important that $\delta $ is not the actual atoms' velocity $p$ but
the effective one $\alpha p$. If the excited fraction $\alpha $ is small
then $\delta $ is considerably smaller than $p$ which in itself can be
larger than unity (i.e., than the speed of sound). Both for $\delta =0$ and $%
p$ independent of density$,$ the soliton with any allowed velocity $v$ and
phase change $\Delta \theta _{sm}$ has an antisoliton with the phase change $%
-\Delta \theta _{sm},$ and the two velocities are symmetric with respect to $%
v=\delta $, Fig.2a. In contrast, Eq.(\ref{slow teta}) shows that a density
dependent $p$ changes the symmetry: not for all $v$ solitons have
antisolitons and the velocities of a soliton and its antisoliton are
asymmetric, Fig.2b.

\begin{figure}[tph]
\includegraphics[width=0.7\textwidth]{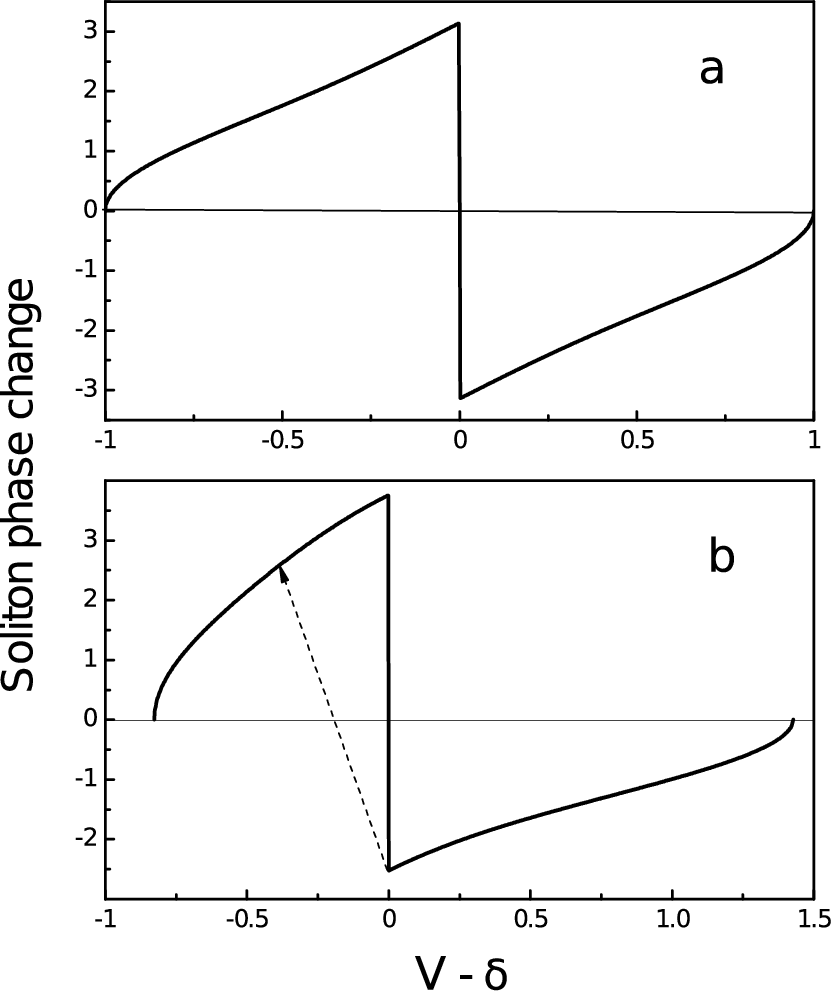}
\caption{Total phase change along the soliton $\Delta \protect\theta _{sm}(v-%
\protect\delta ).$ a) \ $\protect\delta =0$ and $p=const$ independent of
density, every soliton has antisoliton with the opposite phase. b) $p\propto
\left\vert \protect\phi \right\vert ^{2},$ $\protect\delta =0.3,$ solitons
with the phase above the point indicated by the arrow do not have
antisolitons.}
\label{Fig1}
\end{figure}

\begin{figure}[tph]
\centering
\includegraphics[width=0.7\textwidth]{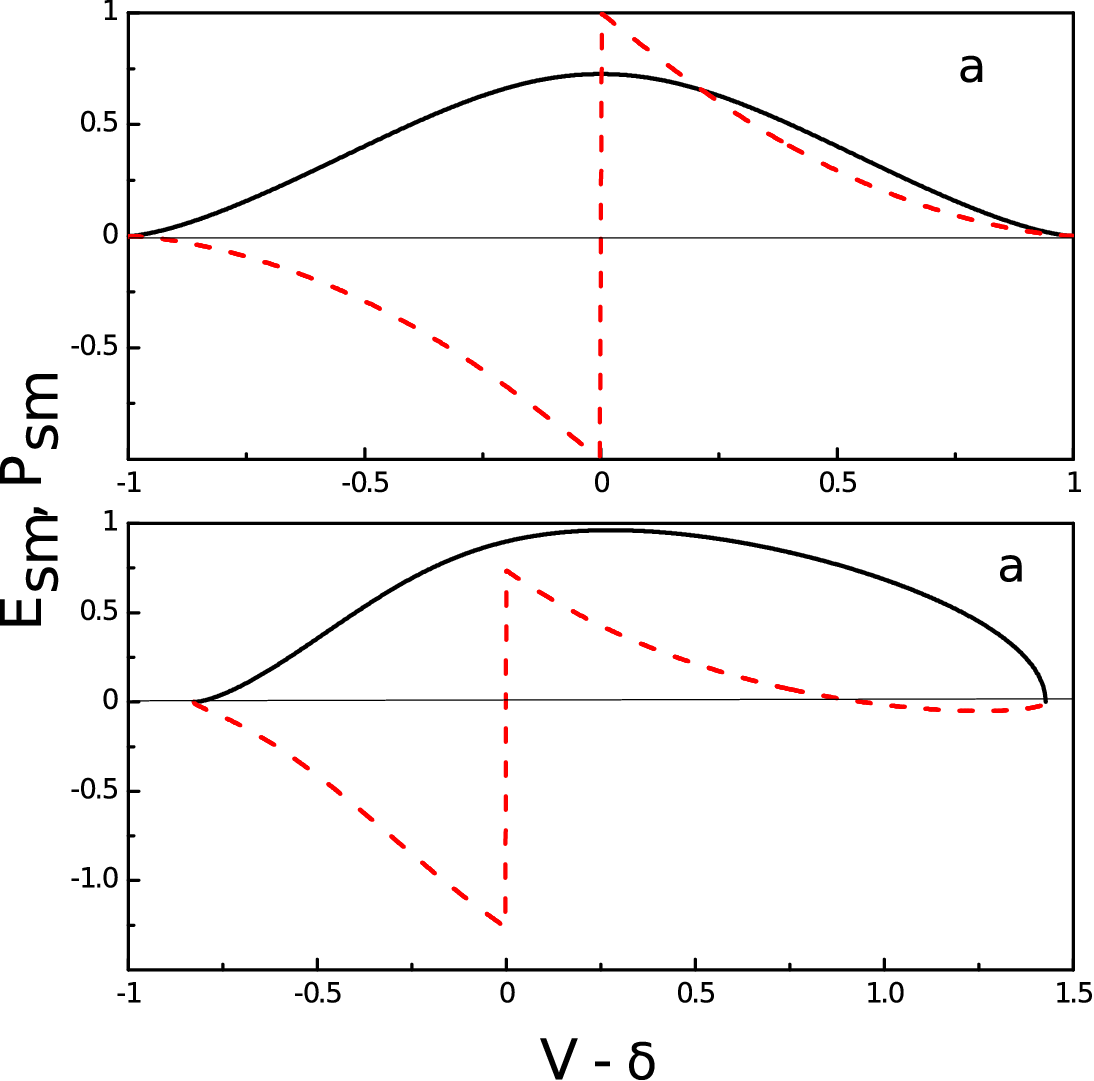}
\caption{Soliton energy $E_{sm}$ (solid lines) and momentum $P_{sm}$ (dashed
lines) vs $(v-\protect\delta ).$ a) $\protect\delta =0$ and $p=const$
independent of density, b) $p\propto \left\vert \protect\phi \right\vert
^{2},$ $\protect\delta =0.3.$ }
\label{Fig2}
\end{figure}

\begin{figure}[tph]
\centering
\includegraphics[width=0.7\textwidth]{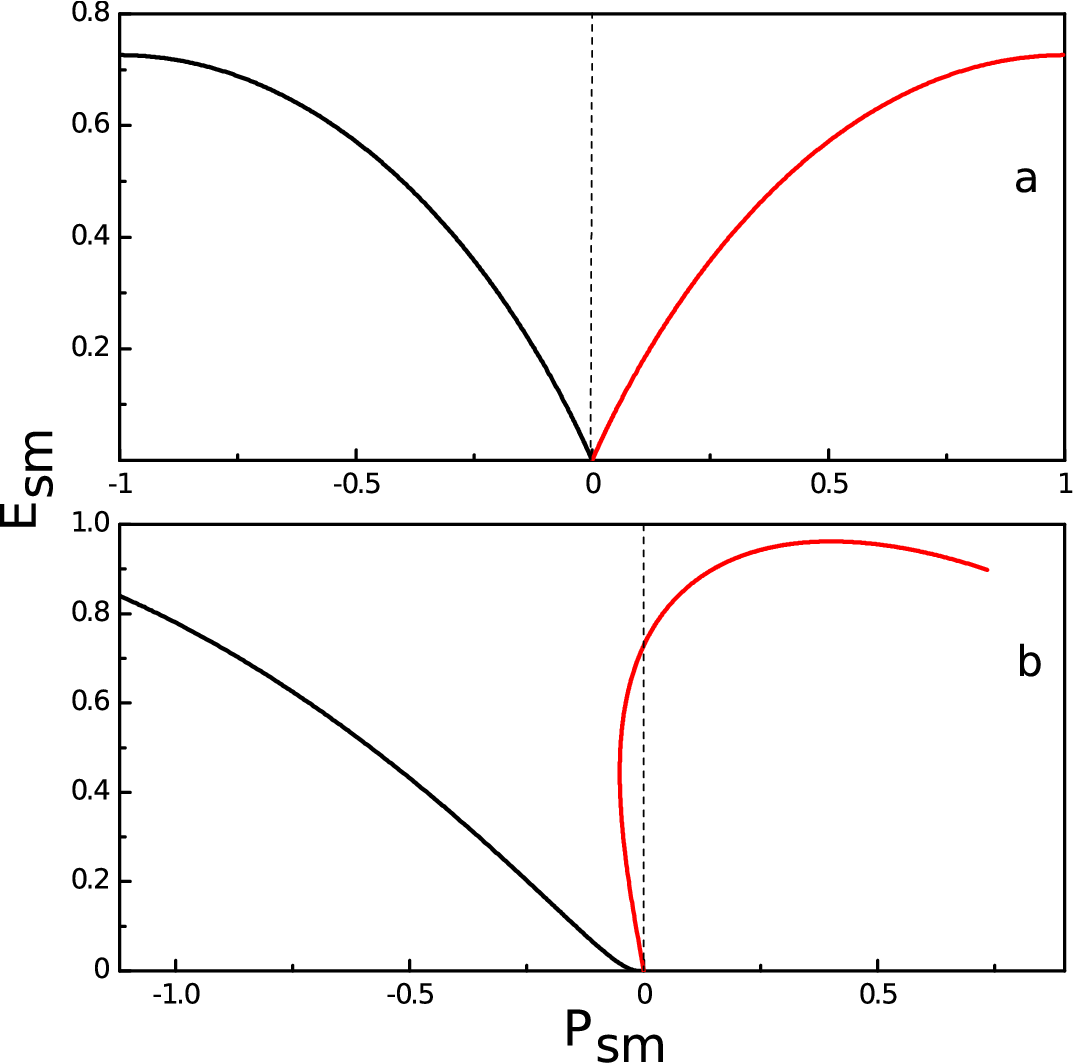}
\caption{Soliton dispersion relation $E_{sm}(P_{sm}).$ a): $\protect\delta %
=0 $ and $p=const$ independent of density. b) $p\propto \left\vert \protect%
\phi \right\vert ^{2},$ $\protect\delta =0.3.$}
\label{Fig4}
\end{figure}

The energy $E_{sm}$ of the slow mode is equal to the total $E$ (\ref{Esol})
minus the contribution from the homogeneous gas with the excitations, see
Appendix C (the excitation energy has to be excluded too, but it was set
zero \cite{Excitation energy}). This $E_{sm}$ and the soliton momentum $%
P_{sm}$ (\ref{psm}) are obtained in the form
\begin{eqnarray}
E_{sm} &=&\frac{\sqrt{3}}{\pi }\left[ 1-(v-\delta )^{2}+G\right] \ln \left[
\frac{\sqrt{3B}+2+3v\delta }{D}\right] +\frac{\sqrt{3}v\delta }{\pi }%
(9\delta ^{2}-1)B,  \label{EEsol} \\
P_{sm} &=&-\frac{\sqrt{3}}{\pi }v(1+3\delta ^{2})\ln \left[ \frac{\sqrt{3B}%
+2+3v\delta }{D}\right] +\frac{\sqrt{3}\delta }{\pi }B-\frac{\Delta \theta
_{sm}}{\pi },  \label{Psm}
\end{eqnarray}%
where $G=4v\delta (1+3v\delta )-2\delta (v-\delta )(1+3v\delta )+3\delta
^{2}\{2/3+v^{2}-(1+3v\delta )^{2}-(v-\delta )^{2}\}$ and the energy is in
units $\hbar ^{2}\pi ^{2}n_{0}^{2}/2m.$ For $\delta =0,$ for which the above
results (\ref{amplitude})-(\ref{Psm}) reproduce those found in \cite{Kolom},
as well as for constant $p$, both $E_{sm}$ and $P_{sm}$ are symmetric with
respect to the value $v=\delta ,$ Fig.3a.\emph{\ }However, for $p\propto
\left\vert \phi \right\vert ^{2},$\ both $E_{sm}(v)$ and $P_{sm}(v)$ are
asymmetric which is shown for $\delta =0.3$ in Fig.3b.

In usual situation, when a soliton is excited in the gas ground state, its
momentum includes only that of a slow mode and the compensating momentum.
Our situation is different in that the total excitation comprises the local
excitations of the momenta $p$ and the slow soliton mode. However, we are
interested in the soliton dispersion relation $E_{sm}$ vs $P_{sm}=P_{tot}-%
\func{mod}(\theta _{tot},2\pi )/\pi $ in which $\theta _{tot}$ contains the
contribution of the momenta $p,$ eq.(\ref{mod2}). Thus, the dispersion
relation $E_{sm}(P_{sm})$ accounts for the soliton excitation in the gas
with local momentum excitations. Fig.4 presents the dependence $%
E_{sm}(P_{sm})$ for $\delta =0$ and a density independent $p,$ and for $%
\delta =0.3$ in the case of $p\propto \left\vert \phi \right\vert ^{2}.$ For
$\delta =0$ and constant $p$ this curve is symmetric with respect to the
line $P_{sm}=0$ \cite{Kolom}, Fig.4a. This is in line with the above rule
that if the momentum $p$ is density independent, then the soliton velocity $%
v $ is just replaced by $v-\delta $ resulting in symmetric $E_{sm}(P_{sm})$.
In our case this momentum is proportional to the density and the dependence $%
E_{sm}(P_{sm})$ is deformed and asymmetric with respect to the line $%
P_{sm}=0,$ Fig.4b$.$ Thus, the reason for the asymmetry of the curves in
Figs.2b, 3b, and 4b as well as for the absence of antisolitons in some
velocity range, Fig.2b, is not a nonzero excitation momentum $p$ itself,\
but its density dependence.

\section{Conclusion}

The development of the physics of ultra cold gases has demonstrated a high
efficiency of the large scale hydrodynamic-type description in terms of
local smoothly varying or coarse-grained quantities. A number of equations
has been presented, e.g., the mean field GPE \cite{Gross,Pitaevskii,Pit},
eGPE with\ the beyond mean field Lee-Huang-Yang corrections \cite{LHY}, the
modified GPE with the quintic nonlinearity \cite{Kolom}, as well as based on
the hydrodynamic analogy MNLSE \cite{Monopole} and LLGPE \cite{Pawl SciPost}%
. All of them have been successfully applied to certain domain of Bose gas
parameters such as dimension, interaction strength, density, and all of them
have somewhat or substantially different derivation and mathematical form
which reflect such parameters. All these derivations have in common that
they have been based on the one-particle quantum description. This motivated
us to derive the general HA to a cold Bose gas which would have a general
form and provide the connection with the many-body quantum mechanical
description. Starting from the many-body quantum mechanical approach we
developed the HA. As any standard HA, it presupposes two different temporal
and spatial scales and, respectively, two different modes, the fast and
slow. Our HA is grounded on the energy functional $E\{\psi _{X},A_{X}\}$,
eqs.(\ref{E}) and (\ref{HX}), of both modes. The fast mode and the local
equilibrium are identified respectively with the many-body wave function $%
\psi _{n}$ and its stationary state at local density $n$. The integration
over the fast mode (over the short scale) naturally resulted in the HE for
the slow mode alone which is the counterpart of the momentum integration in
the local distribution function in the classical HA. The HE contains two
universal terms, the local chemical potential expressed via the energy
eigenvalues of the local $\psi _{n}$ and the interaction between the local
momenta and slow velocity. For different particular forms of the former term
the HE (\ref{SEA}) reduces to the known equations. The second term is new
and its full exploration is a novel problem. In this paper we sketched only
two possible effects related to the local excitations and their density
dependence.

To conclude, we can speculate about possible relevance and applications of
the presence of the momentum related term. We may point to the following
experimental situations and problems. Excitations of local momenta in the
Lieb-Liniger gas have been induced experimentally by a laser beam in \cite%
{Excitations}. States with the peaks at nonzero momenta has been observed in
experiments on colliding clouds of cold Bose atoms. The occupation inversion
obtained in this experiment was attributed to a negative temperature \cite%
{Science,Peotta}. However, the thermodynamic approach to such systems is an
approximation and, in principle, in terms of the HA this situation can be
thought of as an excited state with nonzero momentum.

In the introduction, we mentioned the problem related to the quantum
pressure term, which violates the consistency of the slow mode equation and
results in incorrect interference patterns: on the one hand, it is present
in the exact Schr\"{o}dinger equation and cannot be just discarded, but, on
the other hand, it "spoils" the slow mode equation. The problem is that the
second order spatial density derivative should be better attributed to the
fast mode equation, but the known HAs consist of the equation only for a
slow mode. Our HA, which consists of both slow mode equation (\ref{eXA}) and
fast mode equation (\ref{SEA}), suggests a possible remedy. The quantum
pressure term, which is a part of the term $\triangle A$ in eq.(\ref{SEA}),
can in principle be relocated from this equation to the fast mode equation (%
\ref{eXA}). In this way, the slow mode equation would be more consistent
while the local energy eigenvalue be renormalized. This can hopefully extend
the applicability of the HA to the interference-type effects.

As we saw above, the constant phase of a stationary soliton in a ring
geometry can be compensated by the global flow of the whole gas. However, in
course of a gas dynamics the local phase can rapidly vary and the global
flow is hardly a remedy. Can the local phase be compensated by a local
momenta excitation? Interestingly, such a possibility was mentioned by
Kivshar and coworkers in \cite{Kivshar,Kivshar94}. These papers address gray
solitons of a nonlinear Schr\"{o}dinger equation (similar to the GPE)
describing their propagation in a fast oscillating background. Surprisingly,
though this equation allows only for a constant spatially independent
background wave number, in \cite{Kivshar,Kivshar94} a spatial dependence of
the wave number in the soliton vicinity was mentioned as a possible source
of the phase compensation. Our result seems to be the right tool to approach
this problem.



\section*{Acknowledgements}

Author is grateful to Krzysztof Paw{\l}owski and Kazimierz Rz\c{a}\.{z}ewski
for valuable discussions and advices and to the Center for Theoretical
Physics PAS for hospitality. This research is part of the project No.
2022/45/P/ST3/04237 co-funded by the National Science Centre and the
European Union Framework Programme for Research and Innovation Horizon 2020
under the Marie Sk\l odowska-Curie grant agreement No. 945339.

\appendix{}

\section{Separation of the effect of a slow density variation in the reduced
one-body density matrix.}

Consider a single subsystem $\Delta V_{X}$ with the center at $X$. The
density $n(y)$ is slowly varying within this volume and the psi function is
a functional of $n,$ $\psi (y,x^{N-2}|n)=\psi _{n}(y),$ where for brevity
the dependence on $x$ is omitted. We want to separate the contribution due
to this density variation from the one-body density matrix taken at the
central density $n_{X}=n(X).$ A small change in density $n=n_{X}+\delta n(y)$
results in the change $\psi _{n}(y,x^{N-1})=\psi _{X}(y,x^{N-1})+\delta \psi
(y,x^{N-1})$ of the wave function. Then one has:%
\begin{align}
\rho _{n}(y,y^{\prime })& =\int dx^{N-1}[\psi _{X}^{\ast }(y)+\delta \psi
^{\ast }(y)]\times \lbrack \psi _{X}(y^{\prime })+\delta \psi (y^{\prime })]
\tag{A1} \\
& \simeq \rho _{X}(y,y^{\prime })+\int dx^{N-1}[\psi _{X}^{\ast }(y)\delta
\psi (y^{\prime })+\psi _{X}(y^{\prime })\delta \psi ^{\ast }(y)].  \notag
\end{align}%
Due to the presence of $\delta (y-y^{\prime })$ in the integrand of (\ref{KX}%
), we can consider only the case $\Delta y=y^{\prime }-y\rightarrow 0.$
Then, neglecting the terms on the order $\delta \psi \Delta y,$ the above
one-body density matrix can be reduced to the form%
\begin{equation}
\rho _{n}(y,y^{\prime })=\rho _{X}(y,y^{\prime })+\delta \rho (y^{\prime
})+\delta \rho ^{\ast }(y),  \tag{A2}
\end{equation}%
where $\delta \rho (y^{\prime })=\int dx^{N-1}\psi _{X}^{\ast }(y^{\prime
})\delta \psi (y^{\prime })$ and $\delta \rho ^{\ast }(y)=\int dx^{N-1}\psi
_{X}(y)\delta \psi ^{\ast }(y).$ Since $\rho _{X}(y,y)=f_{X}(y)\neq 0$,
there exists a finite vicinity of $y^{\prime }-y=0$ in which $\rho
_{X}(y,y^{\prime })\neq 0,$ hence one can divide by this function. Then one
has:%
\begin{equation}
\rho _{n}(y,y^{\prime })=\rho _{X}(y,y^{\prime })\left[ 1+\delta \rho
(y^{\prime })/f_{X}+\delta \rho ^{\ast }(y)/f_{X}\right] ,  \tag{A3}
\end{equation}%
where, in the denominators$,$ $\rho _{X}(y,y^{\prime })$ is replaced by $%
f_{X}$ in within the accuracy up to $O(\Delta y^{2})$. Neglecting terms
quadratic in $\delta \rho ,$ this can be cast in the form of eq.(\ref{f=af}%
), i.e.,%
\begin{equation}
\rho _{n}(y,y^{\prime })=\frac{A_{X}(y)A_{X}^{\ast }(y^{\prime })}{f_{X}}%
\rho _{X}(y,y^{\prime }),  \tag{A4}
\end{equation}%
where we introduced the amplitude $A(y)$ and its conjugate according to the
following definition:%
\begin{align}
A_{X}(y)& =\sqrt{f_{X}}+\delta \rho (y)/\sqrt{f_{X}},  \tag{A5} \\
A_{X}^{\ast }(y^{\prime })& =\sqrt{f_{X}}+\delta \rho ^{\ast }(y^{\prime })/%
\sqrt{f_{X}}.  \notag
\end{align}%
For constant density $n_{X}$ eq.(A4) recovers $\rho _{X}(y,y^{\prime })$ and
for $y=y^{\prime }$ it reduces to $f_{X}(y)$. In formulae (A4) and (A5), the
contribution of the density variation within $\Delta V_{X}$ is separated
from the one-body density matrix at $n_{X}$ in the form of the product $%
A_{X}(y)A_{X}^{\ast }(y^{\prime })$ which is employed in the calculation of
the kinetic energy (\ref{KX}).

\section{Coarse-grained correlation function and the long range interaction
energy}

In a liquid state, at large separation $y-y^{\prime }\simeq X-X^{\prime }$,
the pair distribution $G_{2}(y,y^{\prime })$ usually tends to the product $%
f(y)f(y^{\prime })$ of the probability densities, so that in general $%
G_{2}(y,y^{\prime })=f(y)f(y^{\prime })[1+g_{2}(y,y^{\prime })]$ where $%
g_{2} $ is the pair correlation. The coarse-grained hydrodynamic pair
distribution $\rho _{2,XX^{\prime }}$ and coarse-grained correlation $%
g_{2,X,X^{\prime }}$ are defined by the following equation:
\begin{align}
\rho _{2,XX^{\prime }}\left\vert A_{X}\right\vert ^{2}\left\vert
A_{X^{\prime }}\right\vert ^{2}& =\frac{1}{(\Delta V)^{2}}\int_{\Delta
V_{X}}dy\int_{\Delta V_{X^{\prime }}}dy^{\prime }G_{2}(y,y^{\prime })
\tag{B1} \\
& =\left\vert A_{X}\right\vert ^{2}\left\vert A_{X^{\prime }}\right\vert
^{2}(1+g_{2,X,X^{\prime }}),  \notag
\end{align}%
where we made use of the relations (A4), $\rho _{X}(y,y)=$ $f_{X}(y)\simeq
f_{X}.$ Then the energy of the long range interaction is%
\begin{equation}
E_{lr}\{A_{X}\}=\frac{N(N-1)}{2}\int_{V}dy\int_{V}dy^{\prime
}G_{2}(y,y^{\prime })U_{lr}(y-y^{\prime })+N\int_{V}dyU_{ext}(y)\rho (y,y)
\tag{B2}
\end{equation}%
\begin{equation*}
=\frac{N(N-1)}{2}\sum_{\Delta V_{X},\Delta V_{X^{\prime }}}\int_{\Delta
V_{X}}dy\int_{\Delta V_{X^{\prime }}}dy^{\prime }G_{2}(y,y^{\prime
})U_{lr}(y-y^{\prime })+N\sum_{\Delta V_{X}}\int_{\Delta
V_{X}}dyU_{ext}(y)\rho (y,y)
\end{equation*}%
\begin{equation*}
\simeq \frac{N(N-1)}{2}\int dX\int dX^{\prime }\left\vert A_{X}\right\vert
^{2}\left\vert A_{X^{\prime }}\right\vert ^{2}(1+g_{2,XX^{\prime
}})U_{lr}(X-X^{\prime })+N\int dX^{\prime }\left\vert A_{X}\right\vert
^{2}U_{ext}(X-X^{\prime }),
\end{equation*}%
which is eq. (\ref{ELR}).

\section{The soliton energy}

The soliton energy is the difference between the energy $E$ of the system
with soliton and $\left\vert \phi \right\vert ^{2}$ $=f\rightarrow 1$ at the
periphery, and the energy $E_{0}$ of the homogeneous system with the same
number of atoms $N.$As the density $n$ in the area of gray soliton is lower
than $n_{0}=N/L,$ the actual density far from soliton $\widetilde{f}_{\infty
}=\left\vert \widetilde{\phi }_{\infty }\right\vert ^{2}$ is slightly higher
than $f_{\infty }=1,$ so that $\widetilde{f}=f+\delta f$ $\ $where $\delta f$
is nonzero at the periphety and very small in a large system. Then%
\begin{equation}
\int dy\delta f+\int dy(f-1)=0.  \tag{C1}
\end{equation}%
Here we find the soliton energy taking into account this correction at the
periphery. The total energy with $\phi $ satisfying the Euler-Lagrange
equation (\ref{EQ}) can be found in the form of the virial theorem. We
multiply (\ref{EQ}) with $\phi ^{\ast }/\pi $ and integrate over the system
length which, regarding the expression for the energy (\ref{Esol}), gives%
\begin{align}
0& =\frac{1}{\pi }\int_{0}^{L}dy\left( iv\phi ^{\ast }\phi ^{\prime }-\frac{1%
}{2}\phi ^{\ast }\phi ^{\prime \prime }+\frac{1}{2}f^{3}-2ip\phi ^{\ast
}\phi ^{\prime }\left\vert \phi \right\vert ^{2}-\lambda \left\vert \phi
\right\vert ^{2}\right)  \tag{C2} \\
& =E+\frac{1}{\pi }\int_{0}^{L}dy\left( iv\phi ^{\ast }\phi ^{\prime
}-\lambda f+\frac{1}{3}f^{3}-ip\phi ^{\ast }\phi ^{\prime }f\right) ,  \notag
\end{align}%
whence%
\begin{equation}
E=\frac{1}{\pi }\int_{0}^{L}dy\left( -iv\phi ^{\ast }\phi ^{\prime }+\lambda
f-\frac{1}{3}f^{3}+ip\phi ^{\ast }\phi ^{\prime }f\right) .  \tag{C3}
\end{equation}%
Next, from the equation (\ref{EQ}) at the periphery one finds $\lambda
=f_{\infty }^{2}/2=1/2.$ Now we change $f$ to the actual $f+\delta f$ and $%
\lambda $ to $\lambda ^{\prime }=1/2(1+2\delta f_{\infty })$ in the
expression (C3), taking into account that $\int dyf^{k}\delta f=\int
dy\delta f+O($soliton width/$L)$ for any $k$ and that the first and last
terms in (C3) vanish at the periphery. Retaining only terms linear in $%
\delta f$ and making use of (C1), one obtains:%
\begin{equation}
E=\frac{1}{\pi }\int_{0}^{L}dy\left( -iv\phi ^{\ast }\phi ^{\prime }+\frac{1%
}{2}-\frac{1}{3}f^{3}+ip\phi ^{\ast }\phi ^{\prime }f\right) .  \tag{C4}
\end{equation}%
The soliton energy $E_{sm}$ is $E$ minus the contribution from the constant
background: $E_{sm}=E-\int dy(-1^{3}/3+1/2).$ Finally, the renormalized
soliton (slow mode) energy is%
\begin{equation}
E_{sm}=\frac{1}{\pi }\int_{0}^{L}dy\left[ \frac{1}{3}(1-\left\vert \phi
\right\vert ^{6})-iv\phi ^{\ast }\phi ^{\prime }+ip\phi ^{\ast }\phi
^{\prime }\left\vert \phi \right\vert ^{2}\right] ,  \tag{C5}
\end{equation}%
which was computed to give the value (\ref{EEsol}).


\begin{thebibliography}{99}
\bibitem{restrictions} The detailed consideration of the restrictions of the
HA and its intrinsical LDA assumptions is out of scope of this paper.
Nevertheless, we briefly address some of such restrictions which is
necessary to properly present the logic of development of the HA approach to
cold quantum gases.
\bibitem{Review} The number of papers with substantial contributions on the
LDA in a cold Boson gas is so numerous that we decided here to refer only to
a few in which the LDA equations were suggested and their origine was
discussed. However, the review \cite{Dubail} on the generalized
hydrodynamics by Bouchoule and Dubail is simultaneously a very good review
on the achivements and restrictions of the standard LDA to which it contains
many references.
\bibitem{Excitation energy} The excitation energy can be incorporated in the
gas energy. The energies of both excitation branches of the Lieb-Liniger gas
scale as $p^{2}\sim \left\vert \phi \right\vert ^{4}$ and their inclusion
would renormalize the coefficient of the $\left\vert \phi \right\vert ^{6}$
term. However, while such inclusion results in the lengthy and practically
untracktable expressions for the phase, energy, and momentum, the main
effect remains due to the $i\delta $ term. As here we aim to sketch the main
possible effects of the density dependent momenta, the omitting of the
excitation energy seems reasonable. Moreover, as we are interested in the
energy and momentum of a soliton, those of the momentum excitation would
have been substracted from the total quantities anyway.
\end{thebibliography}
\end{document}